\documentclass[fleqn,usenatbib]{mnras}

\usepackage{newtxtext,newtxmath}

\usepackage[T1]{fontenc}
\usepackage{tikz} 
\usetikzlibrary{shapes.geometric, arrows}
\tikzstyle{arrow} = [thick,->,>=stealth]
\usepackage{dirtytalk}
\DeclareRobustCommand{\VAN}[3]{#2}
\let\VANthebibliography\thebibliography
\def\thebibliography{\DeclareRobustCommand{\VAN}[3]{##3}\VANthebibliography}

\usepackage{graphicx}	
\usepackage{amsmath}	
\usepackage{lscape}
\usepackage{booktabs}
\usepackage{longtable}
\usepackage{supertabular}
\usepackage{subcaption}
\usepackage{tablefootnote}

\usepackage{ulem}

\title[Compact discs formed by dead zones]{Compact protoplanetary discs can be produced by dead zones}

\author[Tong \& Alexander]{
Simin Tong\thanks{E-mail: st547@leicester.ac.uk, astro.stong@gmail.com}
and Richard Alexander
\\
School of Physics \& Astronomy, University of Leicester, University Road, Leicester, LE1 7RH, UK\\
}

\date{Accepted XXX. Received YYY; in original form ZZZ}

\pubyear{2024}

\begin{document}
\label{firstpage}
\pagerange{\pageref{firstpage}--\pageref{lastpage}}
\maketitle

\begin{abstract}

Radially compact protoplanetary discs ($\lesssim$\,$50\,\mathrm{au}$) are ubiquitous in nearby star-forming regions. Multiple mechanisms have been invoked to interpret various compact discs. In this paper, we propose that fragmentation of fragile dust grains in moderate turbulence, as expected beyond the dead zone, provides an effective alternative mechanism to form compact discs which are consistent with current observations. We run 1-D dust transport and collision models with {\sc DustPy} and generate synthetic observations, and find that discs formed by this mechanism have sizes determined by the extent of their dead zones. Accounting for dust porosity, and considering less fragile dust, do not change disc sizes significantly. The smooth dust morphology can be altered only when pressure bumps are present in the dead zone. However, when present at small radii ($\lesssim$\,$10$\,au), pressure bumps cannot effectively trap dust. Dust in these bumps fragments and replenishes the inner discs, effectively hiding dust traps in the optically thick inner disc from observations. We note a striking resemblance in the radial intensity profile between our synthetic observations and some recent high-resolution observations of compact discs. We discuss how such observations can inform our understanding of the underlying disc physics.
\end{abstract}

\begin{keywords}
accretion, accretion discs -- protoplanetary discs -- stars: pre-main-sequence
\end{keywords}




\section{Introduction}
Protoplanetary discs form as a by-product of angular momentum conservation during star formation, and are the sites of planet formation. In the last decade, the Atacama Large Millimetre/submillimetre Array (ALMA) has surveyed hundreds of proptoplanetary discs in nearby star-forming regions \citep[e.g.][]{2016ApJ...828...46A, 2016ApJ...831..125P, 2016ApJ...827..142B, 2017AJ....153..240A, 2019MNRAS.482..698C, 2019A&A...626A..11C, 2021A&A...653A..46V, 2022ApJ93855A}. Radially extended discs  ($\gtrsim$\,$50~\mathrm{au}$) revealed in these surveys show various substructures, including rings (gaps) \citep[e.g.][]{2018ApJ...869L..42H}, cavities \citep[e.g.][]{2018ApJ...859...32P}, spiral arms \citep[e.g.][]{2018ApJ...860..124D, 2024Natur.633...58S} and crescent-shaped features \citep[e.g.][]{2013Sci...340.1199V}. Formation of these substructures is usually associated with disc-planet interactions \citep[e.g.][]{2012ARA&A..50..211K}, (magneto-)hydrodynamic (MHD) instabilities \citep[e.g.][]{2023ASPC..534..465L} and other physico-chemical effects \citep[e.g.][]{2015ApJ...806L...7Z, 2016ApJ...821...82O}. 

However, smaller ($\lesssim 50~\mathrm{au}$) and seemingly featureless discs, which are prevalent but probably not (well-)resolved in those star-forming region surveys have not been as well studied. Questions remain as to their true radial extent \citep[e.g.][]{2022MNRAS.515L..23I} and whether they also include substructures that are not resolved. Recently, higher-resolution observations and visibility modelling of a few systems suggest that compact discs can also be substructured, and have low contrast gap-ring or `shoulder' features \citep[e.g.][]{2021A&A...645A.139K, 2021ApJ...923..121Y, 2022MNRAS.514.6053J, 2023ApJ...952..108Z, 2024A&A...682A..55M, 2024ApJ...966...59S, 2024PASJ...76..437Y, 2024arXiv241003823H}. These substructures are required in the early stages to match the observed spectral indices \citep{2024A&A...688A..81D} and retain sufficient dust \citep{2022A&A...668A.104S}. They have been detected in some young systems \citep[e.g.][]{2020Natur.586..228S, 2024A&A...689L...5M}, but may remain invisible in some cases \citep{2024arXiv241009042N}. 

\par On the contrary, some discs show no evidence of substructures in the image and visibility planes. One possibility is that radial drift dominates dust transport if the gas-to-dust size ratio is large \citep[e.g.][]{2019A&A...626L...2F, 2019ApJ...882...49L, 2021A&A...645A.139K}. These discs lack dust traps that can effectively prevent the rapid dust loss \citep[e.g.][]{1972fpp..conf..211W, 1977MNRAS.180...57W, 2012A&A...538A...114P}. An alternative is that discs are born small, with less gas and dust than radially extended discs, if the size ratio is moderate \citep[e.g.][]{2024A&A...682A..55M}. The tendency for smooth discs to be smaller \citep[e.g.][]{2019ApJ...882...49L}, less luminous \citep[e.g.][]{2017ApJ...845...44T, 2018ApJ...865..157A, 2023ApJ...952..108Z}, and hence less massive \citep[assuming low optical depth, e.g.][]{2021AJ....162...28V} than substructured discs supports both interpretations. However, it is hard to infer whether substructures exist in seemingly smooth discs but remain undetected due to observational effects -- such as insufficient resolution, the large optical depth \citep[][]{2023A&A...673A..77R}, high inclination \citep{2022arXiv220408225V} -- or they are genuinely absent. \citet{2019ApJ...882...49L} also raise the possibility that rings are present further out but remain undetected.

From a theoretical perspective, understanding dust evolution requires knowledge of the gas. The evolution of gas can be driven by several mechanisms, including magneto-rotational instabilities \citep[MRI,][]{1991ApJ...376..214B, 1995ApJ...440..742H}, magneto-hydrodynamical winds \citep[e.g.][]{1982MNRAS.199..883B, 2013ApJ...769...76B}, and other hydrodynamical instabilities, \citep[such as vertical shear instabilities;][]{2004A&A...426..755A,2013MNRAS.435.2610N}. The MRI has long been thought as the main driver of disc turbulence, and its efficiency is usually parametrized in terms of the \citet{1973A&A....24..337S} parameter $\alpha_\mathrm{SS}$. Its operation depends on the ionization chemistry and configuration of magnetic fields. In the region sandwiched between the very inner ($\lesssim 1~\mathrm{au}$) \citep[e.g.][]{2008MNRAS.388.1223S, 2016ApJ...827..144F, 2019A&A...630A.147F} and the outer disc ($\gtrsim$ tens of $\mathrm{au}$) \cite[e.g.][]{2013ApJ...769...76B,2015ApJ...798...84B, 2023ASPC..534..465L}, the gas near the disc midplane is not well ionised, and the disc becomes MRI-quenched, leading to a low vertically averaged $\alpha_\mathrm{SS}$. 
This results in a disc where $\alpha_\mathrm{SS}(R)$ becomes a varying function of radius, and the MRI-inactive region is referred to as the dead zone \citep{1996ApJ...457..355G}. The low turbulence environment in the dead zone is thought to be favourable for pebble and planetesimal formation, but observational evidence indicating the existence of dead zones is limited \citep[e.g.][]{2019ApJ...886..103O, 2022ApJ...930...56U}. 

In this paper we propose an alternative way to form radially compact discs in the fragmentation-dominated regime, where the maximum grain size is limited by fragmentation. This new scenario requires the presence of the dead zone, beyond which turbulence is too high to grow mm-dust when the dust is fragile \citep[as suggested by recent laboratory experiments, e.g.][]{2019ApJ...873...58M}. We run 1-D dust evolution models using \textsc{DustPy} \citep{2022ApJ...935...35S} to explore various dead zone models, including sharp and slow transitions, and examine how the location of planet-carved gaps affects dust morphologies. The paper is structured as follows: we present our models and their setup in Section \ref{sec:method} and \ref{sec:setup}, respectively. We show the main results in Section \ref{sec:result}, and discuss these results and their observational implications in Section \ref{sec:discussion}. We summarize our conclusions in Section \ref{sec:conclusion}. 

\section{Method}\label{sec:method}
\subsection{Gas disc evolution}
We consider geometrically thin, axisymmetrical protoplanetary discs driven by $\alpha$-parametrized viscosity \citep[$\alpha_\mathrm{SS}$,][]{1973A&A....24..337S, 1974MNRAS.168..603L}

\begin{equation}\label{eq:master}
    \frac{\partial \Sigma_g}{\partial t}= \frac{3}{R}\frac{\partial}{\partial R}\Biggl[R^{1/2}\frac{\partial }{\partial R
    }(\nu_\mathrm{SS} \Sigma_g R^{1/2})\Biggr].
 \end{equation}
Here $\Sigma_g$ is the gas surface density.  $\nu_\mathrm{SS}=\alpha_\mathrm{SS}c_sH_g$ is the kinematic viscosity. $c_s=\sqrt{k_BT/(\mu m_H)}$ is the speed of sound and $H_g$ is the gas scale-height. $k_B$ is the Boltzmann constant, and $T$ is the mid-plane temperature. $m_H$ is the mass of atomic hydrogen and $\mu=2.3$ is the mean molecular mass in units of $m_H$. 
$\Omega_K=\sqrt{GM_*/R^3}$ is the Keplerian angular velocity orbiting a central star of $1~M_*$ at the radius $R$. $G$ is the gravitational constant. 

We adopt a time-independent disc temperature, which has reached thermal equilibrium between heating and cooling. It obeys 
\begin{equation}\label{eq:thermal}
	T(r) = \Bigl(\frac{\varphi L_*}{8\pi r^2\sigma_\mathrm{SB}}\Bigr)^{1/4},
\end{equation}
where $\varphi$ is the irradiation angle and $\sigma_\mathrm{SB}$ is the Sefan-Boltzmann constant.  The stellar luminosity, $L_*$, is consistently computed from stellar parameters: the stellar radius $R_*$ and the effective temperature $T_\mathrm{eff}$ by $L_* = 4\pi R_*^2 \sigma_\mathrm{SB}T_\mathrm{eff}^4$.

\subsection{Dust disc evolution}\label{sec:method_dust}
We follow the dust framework built in \textsc{DustPy} \citep{2022ApJ...935...35S}, which is a dust evolution model incorporating dust transport and dust collisions. The latter can be either constructive or destructive. Coagulation occurs when the relative colliding velocity $\Delta v$ is below a certain threshold, known as the fragmentation velocity -- a free parameter in the model. At relative velocities well above this threshold collisions result in fragmentation and erosion, replenishing smaller particles. Bouncing is not considered in \textsc{DustPy} by default. Coagulation and fragmentation are computed by solving the Smoluchowski equation \citep{1916ZPhy...17..557S}.

The dust follows the advection-diffusion equation \citep{2010A&A...513A..79B}
\begin{equation}
	\frac{\partial \Sigma_d^i}{\partial t}+\frac{1}{R}\frac{\partial}{\partial R}\Big(R\Sigma_d^i v_{r,d}^i\Big)- \frac{1}{R}\frac{\partial}{\partial R}\Bigg[RD_d^i\Sigma_g \frac{\partial}{\partial R}\Big(\frac{\Sigma_d^i}{\Sigma_g}\Big)\Bigg]=0.
\end{equation}
where $\Sigma_d^i$ and $v_{r,d}^i$ are the surface density and the radial velocity (Section \ref{sec:method_dust_vr}) of species $i$, respectively. $D_d^i$ is the dust diffusivity of species $i$, describing the radial diffusion of dust due to turbulence \citep{2007Icar..192..588Y} 
\begin{equation}\label{eq:diff}
	D_d^i = \frac{\delta_r c_s^2}{\Omega_K}\frac{1}{1+\mathrm{St}_i^2 }.
\end{equation}
$\delta_r$ is the radial dust diffusion parameter, and by default has a value equal to the gas transport parameter $\alpha_\mathrm{SS}$. Here gas drag falls into the Epstein regime, where the gas mean free path $\lambda_\mathrm{mfp}$ is larger than the dust particle size $a_i$ ($\lambda_\mathrm{mfp}\geq 4a_i/9$), so the Stokes number $\mathrm{St}_{i}$ is defined as
\begin{equation}\label{eq:stokes}
	\mathrm{St}_i= \frac{\pi}{2}\frac{a_i\rho_s}{\Sigma_g},
\end{equation}
where $\rho_s$ is the dust monomer (material) density.

Dust mass and hence sizes are discretized in \textsc{DustyPy}. This makes $\Sigma_d$ dependent on the assumed mass grid. A grid-independent quantity $\sigma_d$ \citep{2010A&A...513A..79B}
\begin{equation}
	\Sigma_d(r) = \int_0^{\infty} \sigma_d(r, a) d\ln a
\end{equation} 
is therefore introduced as an alternative surface density. 

\subsubsection{Dust radial velocity}\label{sec:method_dust_vr}
Gas follows sub-Keplerian motion due to its experience of an outward pressure that dust does not feel. Small dust particles are well coupled with very short stopping times, so move with the gas. Large particles move faster than gas and hence feel `headwind' drag, which causes them to lose angular momentum and spiral inwards. The radial drift of dust is determined by its Stokes number, and can be written as \citep{1986Icar...67..375N, 2002ApJ...581.1344T}
\begin{equation}\label{eq:dust_vr}
	v_{r,d}^{i} = \frac{1}{1+\mathrm{St}_{i}^2}v_g-\frac{2\mathrm{St}_{i}}{1+\mathrm{St}_{i}^2}\eta v_k,
\end{equation}
where
\begin{equation}\label{eq:dust_eta}
	\eta = -\frac{1}{2}\Bigl(\frac{H_g}{R}\Bigr)^2\frac{\partial \ln P}{\partial \ln R},
\end{equation}
$P=\rho_\mathrm{mid} c_s^2$ is the gas pressure at the midplane (assuming the disc is vertically isothermal), and $v_k= \Omega_K R$ is the Keplerian velocity. When $\mathrm{St}_i\ll 1$, we have $v_{r,d}^{i}\sim v_g$, and the second term in Eq. \ref{eq:dust_vr} is negligible. The second term -- which drives radial drift -- peaks when $\mathrm{St}_{i} \sim 1$.
  
\subsubsection{Vertical Settling}\label{sec:method_dust_vs}
Dust can be stirred up by turbulence. When turbulence mixing and gravitational settling reach an equilibrium, the scale-height $H^{i}_{d}$ of dust species $i$ in an isothermal gas disc is \citep{1995Icar..114..237D}
\begin{equation}\label{eq:dust_sh}
	H_{d}^i=H_g\sqrt{\frac{\delta_\mathrm{vert}}{\mathrm{St}_i+\delta_\mathrm{vert}}},
\end{equation}
where $\delta_\mathrm{vert}$ parametrizes the efficiency of turbulent mixing in the vertical direction, and it is taken as equivalent to gas transport $\alpha_\mathrm{SS}$ by default. The dust vertical distribution is assumed to be Gaussian 
\begin{equation}
	\rho_{d}^i(z) = \frac{\Sigma_{d}^i}{\sqrt{2\pi}H_d^i}\exp \Bigg(-\frac{-z^2}{{H^i_d}^2}\Bigg)
\end{equation}
to allow analytic solutions \citep{2010A&A...513A..79B}. In our vertically averaged 1-D models the vertical structure is applied to compute the radiative transfer models, which are further used to generate synthetic observations. 

\subsubsection{Dust velocity}
The dust velocity is composed of several components, including the radial velocity (see Section \ref{sec:method_dust_vr}), the azimuthal velocity, the velocity from vertical settling (see Section \ref{sec:method_dust_vs}), and velocities from turbulent \citep{2007A&A...466..413O} and Brownian motion. All of these depend on the particle size. The total velocity is a quadratic sum of these components. The relative velocity $\Delta v$ between particles follows the Maxwell-Boltzmann distribution \citep{2012A&A...544L..16W}, and the outcome of collisions is determined by the magnitude of $\Delta v$ relative to the fragmentation velocity $v_\mathrm{frag}$.
 
\subsubsection{Fragmentation- and drift-limited regimes}
The maximum particle size can be limited either by radial drift (i.e. particles drift faster than they grow), or by fragmentation (i.e. collisions between particles exceeding a certain size result in fragmentation rather than growth). This defines two regimes for dust discs.

When the fragmentation velocity is relatively low, the entire disc (or at least a large fraction of it) is in the fragmentation-limited regime \citep[e.g.][]{2009A&A...503L...5B, 2024A&A...682A..32J}. For a non-laminar disc, the turbulent motion typically dominates the relative velocity. By equating the relative velocity between turbulent motion of similar sized particles given in \citet{2007A&A...466..413O} to the fragementation velocity, the fragmentation-limited Stokes number and particle size in the Epstein regime are \citep{2009A&A...503L...5B, 2012A&A...539A.148B}
\begin{equation}\label{eq:st_frag}
 	\mathrm{St_{frag}}= \frac{1}{3\delta_\mathrm{turb}}\Big(\frac{v_\mathrm{frag}}{c_s}\Big)^2.
\end{equation}
and
\begin{equation}\label{eq:a_frag}
    a_\mathrm{frag}= \frac{2}{3\pi}\frac{\Sigma_g}{\rho_s \delta_\mathrm{turb}}\Big(\frac{v_\mathrm{frag}}{c_s}\Big)^2,
\end{equation}
respectively. When the disc is laminar ($\delta_\mathrm{turb}=\alpha_\mathrm{SS}\lesssim 10^{-5}$), or dust drift induced by the pressure gradient dominates (see Eq. \ref{eq:dust_vr} and \ref{eq:dust_eta}), the major contributor to the relative velocity becomes radial drift. The limiting Stokes number and particle size are hence given by balancing the relative radial drift and the fragmentation velocity \citep{2012A&A...539A.148B}
\begin{equation}\label{eq:st_df}
	\mathrm{St_{df}}=\frac{v_\mathrm{frag} v_k}{c_s^2(1-N)}\bigg| \frac{d\ln P}{d\ln R}\bigg|^{-1} 
\end{equation}
and 
\begin{equation}\label{eq:a_df}
	a_\mathrm{df}= \frac{2}{\pi} \frac{\Sigma_g}{\rho_s (1-N)}\frac{v_\mathrm{frag}v_k}{c_s^2}\bigg| \frac{d\ln P}{d\ln R}\bigg|^{-1},
\end{equation}
where $N$ ($<1$) is the ratio of Stokes numbers between the smaller and larger collision particles, and is reasonably assumed to be $0.5$ \citep{2012A&A...539A.148B}. This is the so-called drift-fragmentation limit.

By contrast, when the dust drifts inwards before it can grow, we have the drift-limited regime, where the limiting particle size is obtained by equating the growth timescale to the drift timescale. The Stokes number and its corresponding particle size are \citep{2012A&A...539A.148B} 
\begin{equation}\label{eq:st_drift}
	\mathrm{St}_\mathrm{drift} = \frac{\Sigma_d}{\Sigma_g}\bigg(\frac{v_k}{c_s}\bigg)^2\bigg| \frac{d\ln P}{d\ln R}\bigg|^{-1},
\end{equation}
and
\begin{equation}\label{eq:a_drift}
    a_\mathrm{drift}=\frac{2\Sigma_d}{\pi \rho_s}\bigg(\frac{v_k}{c_s}\bigg)^2\bigg| \frac{d\ln P}{d\ln R}\bigg|^{-1}.
\end{equation}

Eq. \ref{eq:st_frag}, \ref{eq:st_df} and \ref{eq:st_drift} (or Eq. \ref{eq:a_frag}, \ref{eq:a_df} and \ref{eq:a_drift}) can be used to estimate the maximum particle size at a given radius.

\subsection{Dead zone model}\label{sec:method_DZ}
The physics governing the formation of dead zones is complex, but is usually approximated by a simple prescription in 1-D models \cite[e.g.,][]{2012MNRAS.420.2851M, 2016A&A...596A..81P, 2021A&A...655A..18G, 2024_disc_evo}. 
In this work, we adopt a transition profile
\begin{equation}\label{eq:alpha_ss}
    \alpha_\mathrm{SS} = \frac{\alpha_\mathrm{SS,MRI}-\alpha_\mathrm{SS,DZ}}{1 + \exp{[-w\cdot(R-R_t)/R]}} + \alpha_\mathrm{SS,DZ}
\end{equation}
where $\alpha_\mathrm{SS,DZ}$ and $\alpha_\mathrm{SS,MRI}$ are $\alpha_\mathrm{SS}$ in the MRI dead zone and MRI-active region, respectively. $R_\mathrm{t}$ is the transition radius, beyond which the disc is in the fully MRI-active regime. $w$ is the parameter tuning the width of the transition, which is taken to be $w=30$ for a sharp (but continuous and differentiable) transition. A slow transition profile, mimicking results from non-ideal MHD simulations that account for all the three non-ideal MHD effects \citep{2016ApJ...818..152B}, is also studied in Section \ref{sec:result}.

\subsection{Dust trap model}\label{sec:method_dusttrap}
Pressure bumps act as dust traps, and are thought to be necessary to retain sufficient dust for millions of years \citep[e.g.][]{1972fpp..conf..211W, 2012A&A...538A...114P}. They are required to match numerous observables, and are also invoked to explain the dust rings prevalent in ALMA observations \citep[e.g.][]{2018ApJ...869L..46D}. 

One of the most popular mechanisms to form a dust trap is disc-planet interactions. Planets perturb the gas, and can carve a gap if they are sufficiently massive. The change in the gas density creates a pressure bump, which acts as a dust trap and retains dust locally. We mimic this effect by varying $\alpha_\mathrm{SS}$, following \citet{2018ApJ...869L..46D}. The perturbed $\alpha_\mathrm{pert}$ is prescribed by 
\begin{equation}\label{eq:bump}
	\alpha_\mathrm{pert} = \alpha_\mathrm{SS}\cdot \Bigg[1+ A_\mathrm{gap}\cdot \exp \Big(-\frac{(R-R_\mathrm{gap})^2}{2w_\mathrm{gap}^2} \Big) \Bigg],
\end{equation}
where $A_\mathrm{gap}$, $R_\mathrm{gap}$ and $w_\mathrm{gap}$ are the amplitude, location and width of the Gaussian bump in $\alpha_\mathrm{SS}$, respectively. By default we assume $w_\mathrm{gap}=2H_g$, to maintain the stability of the bump \citep[e.g.][]{2000ApJ...533.1023L, 2009A&A...497..869L}. In our models we assume that bumps are present at the start of the simulation ($t=0$), which corresponds to the start of the Class II phase (when infall from the protostellar envelope is negligible, and the disc is gravitationally stable).
The perturbed and unperturbed gas surface densities, and other disc properties (such as the gas scale-height and $\alpha_\mathrm{SS}$), are then used to estimate the planetary mass, following \citet{2018ApJ...861..140K}.

\subsection{Radiative transfer and synthetic observations}
Post-processing radiative transfer for {\sc DustPy} models is implemented using the integrated {\sc RADMC-3D} \citep{2012ascl.soft02015D}\footnote{\url{https://www.ita.uni-heidelberg.de/~dullemond/software/radmc-3d/index.php}}  in {\sc DustPyLib} \citep{2023ascl.soft10005S} and assuming {\sc DSHARP} dust opacites \citep{2018ApJ...869L..45B}\footnote{\url{https://github.com/birnstiel/dsharp_opac/}}. The dust is assumed to be compact and composed of $36.42\%$ (volume fraction) water ice \citep{2008JGRD..11314220W}, $16.7\%$ astronomical silicates \citep{2003ARA&A..41..241D}, $2.58\%$ troilite and $44.30\%$ refractory organics \citep{1996A&A...311..291H}. The radiative transfer models are computed at ALMA Band 3 ($3~\mathrm{mm}$) and Band 6 ($1.3~\mathrm{mm}$) assuming an inclination of $0^{\circ}$ (face-on discs). They are further used for generating synthetic observations  with {\sc CASA} \citep{2022PASP..134k4501C}. The beam size for both bands is approximately $0^{\prime \prime}05$ with an integration time of $30~\mathrm{mins}$, unless specified otherwise.

Radiative transfer is computed in spherical coordinates, with $10^7$ photons for thermal radiation and $10^6$ photons for anisotropic scattering. The radial grid is directly adopted from {\sc DustPy} models (500 cells covering $1-2500~\mathrm{au}$), and default setups are assumed for the azimuthal and polar grids. The former is a coarse grid of $16$ cells from $0$ to $2\pi$, while the latter has $256$ cells from $0$ to $\pi$. The particle grid has 50 logarithmically-spaced bins from the smallest grain size in each model to either $10~\mathrm{cm}$ or the maximum grain size in the model, whichever is larger. The central star is treated as a blackbody, with parameters adopted from {\sc DustPy} models (see Table \ref{tb:params}). 

{\sc CASA} generates both noise-free and thermal noise-included visibility measurements. We use the latter to evaluate the noise level, and assess the observability of faint features seen in the noise-free images. All the synthetic observations shown in this work are noise-free images, from which we also measure the continuum fluxes.

\section{Simulation Setup}\label{sec:setup}
The radial grid has $500$ cells logarithmically distributed from $1~\mathrm{au}$ to $2500~\mathrm{au}$. This ensures that all the dust rings (see Section \ref{sec:method_dusttrap}) are numerically resolved, with the full width at half maximum of each ring resolved into at least 4 grid cells. Our initial condition for the gas surface density follows the self-similar solution
\begin{equation}
	\Sigma_g(R) = \frac{M_{g}}{2\pi R_{c}^2} \Bigl(\frac{R}{R_{c}}\Bigr)^{-1}\exp{\Bigl(-\frac{R}{R_{c}}\Bigr)}
\end{equation}
where $M_{g}$ is the initial gas disc mass and $R_{c}$ is the initial characteristic radius. We adopt a dust-to-gas mass ratio of $0.01$ as the standard interstellar medium value\footnote{This is principally equivalent to setting dust-to-gas surface density ratio, which is a parameter in \textsc{DustPy}, to be $0.01$. However, dust with large Stokes numbers in the outer disc, which can drift inwards rapidly, is removed automatically in \textsc{DustPy}. Therefore, we adopt a corrected dust-to-gas surface density that is slightly higher than $0.01$, to ensure that sufficient dust is included in the initial disc.}. However, such initial profiles can cause unwanted artefacts in numerical simulations when a dead zone model, i.e. radially varying $\alpha_\mathrm{SS}$, is included (see also Section \ref{sec:result_DZ_revis}). We therefore allow the gas surface density to relax before adding the dust. The initial dust surface density is hence similar to the relaxed gas surface density.

The particle grid ranges $0.5~\mu \mathrm{m}$--$24~\mathrm{cm}$ in $120$ mass bins, corresponding to $10^{-12}$--$10^{5}~\mathrm{g}$ when the monomer (material) density is $\rho_s = 1.67~\mathrm{g~cm^{-3}}$. Seven bins are employed per decade of dust mass, to ensure accuracy and computational feasibility \citep{2014A&A...567A..38D}. The initial dust size distribution follows the MRN distribution $n(a)\propto a^{-3.5}$ \citep{1977ApJ...217..425M}. The maximum initial size is set to be either $1~\mu \mathrm{m}$ or a `safe' size that does not experience strong inwards drift, whichever is smaller. Particles in all the models shown in Section \ref{sec:result} are assumed to be compact and spherical. 

The temperature varies with the radius as $T\propto r^{-1/2}$ (see Eq. \ref{eq:thermal}). We assume the central star to be a typical T Tauri star, with stellar radius $R_* = 2~\mathrm{R_\odot}$ and stellar mass $M_* = 1~\mathrm{M_\odot}$. The effective temperature is $T_\mathrm{eff}= 3500~\mathrm{K}$. The irradiation angle $\varphi$ in Eq. \ref{eq:thermal} is fixed to be $0.05$ in our models. The thermal structure is hence equivalent to $T= 134~\mathrm{K}(R/\mathrm{au})^{-1/2}$.

Three dust parameters are employed to describe dust radial diffusion $\delta_\mathrm{rad}$, dust turbulent mixing in the radial direction $\delta_\mathrm{turb}$ and in the vertical direction $\delta_\mathrm{vert}$. They have the same value as the gas transport parameter $\alpha_\mathrm{SS}$. We keep all the three $\delta$ parameters unchanged when $\alpha_\mathrm{SS}$ is varied to mimic the gaps opened by planets in gas discs (see Section \ref{sec:method_dusttrap}). This means that the radial diffusion, scale-height, and turbulent motion of the dust are not affected by the introduction of dust bumps.

Ice grains, which were thought to be sticky when colliding at a high relative velocity ($\sim10~\mathrm{m~s^{-1}}$) and therefore favour dust growth \citep[e.g.][]{2008ARA&A..46...21B, 2015ApJ...798...34G}, were recently found to have no advantages over silicates \citep[e.g.][]{2019ApJ...873...58M}, whose fragmentation velocity is lower ($\sim 1~\mathrm{m~s^{-1}}$). \citet{2024NatAs...8.1148U} also find that fragile dust is required to explain multi-wavelength and polarization observations. In light of this, we set $v_\mathrm{frag}=1~\mathrm{m~s^{-1}}$ as the canonical fragmentation threshold in our models. When a higher $v_\mathrm{frag}$ is considered, we lift the maximum in the dust mass grid from $10^{5}$ to $10^{7}~\mathrm{g}$, i.e. $a\sim 0.5~\mu \mathrm{m}$--$110~\mathrm{cm}$. 

The transition from MRI-active to MRI-inactive regions is discussed in detail in Section \ref{sec:result_formation}. We assume that viscosity in the dead zone is purely from hydrodynamic instabilities and has a value of $\alpha_\mathrm{SS,DZ}=10^{-4}$. We vary $\alpha_\mathrm{SS,MRI}$ in the MRI-active region from $10^{-2}$ to $5\times10^{-4}$, and set the transition radius at $R_t=20$, $30$ and $50~\mathrm{au}$. 

The parameters for our \textsc{DustPy} models and dead zone models are listed in Table \ref{tb:params}. The discs considered in this work are all around isolated single stars. All the models in this work are computed with \textsc{DustPy} version 1.0.4.

\begin{table}
\caption{Input parameters for our disc models. Where more than one value of a parameter is used, the canonical value is highlighted in {\bf bold}.}\label{tb:params}
\begin{tabular}{ccc}
\hline
\hline
Parameters & Symbols [units] & Values\\   
\hline
Stellar mass  & $M_*$ [$M_\odot$] & 1 \\
Stellar radius & $R_*$ [$R_\odot$]  & 2 \\
Effective stellar temperature   & $T_\mathrm{eff}$ [$\mathrm{K}$] & 3500\\
Initial gas disc mass & $M_g$ [$M_\odot$]                     & 0.05\\
Initial dust-to-gas mass ratio & $\epsilon$                                 & 0.01 \\
Characteristic radius $^{*}$& $R_c$ [$\mathrm{au}$] & 60\\
Turbulence in dead zones  & $\alpha_\mathrm{SS, DZ}$ & $10^{-4}$\\
Turbulence in MRI-active zones  & $\alpha_\mathrm{SS, MRI}$ & $10^{-2}$, $\mathbf{10^{-3}}$, $5\times 10^{-4}$ \\
Transition radius  & $R_t$ [au] & $20$, $\mathbf{30}$, $50$ \\
Fragmentation velocity  & $v_\mathrm{frag}$  [$\mathrm{m~s^{-1}}$] & $1$    \\
Dust monomer density & $\rho_s$ [$\mathrm{g~cm^{-3}}$]  & 1.67 \\ 
Distance to the disc & $d$ [$\mathrm{pc}$] & 150 \\
\hline
\end{tabular}
\footnotesize{\textbf{Note}: $^{*}$The characteristic radius shown here is the radius before the gas surface density is relaxed.}\\
\end{table}

\section{Results}\label{sec:result}
\subsection{Revisiting dead zones}\label{sec:result_DZ_revis}

Previous studies have found that pressure bumps formed at dead zone outer edges can effectively trap dust, potentially explaining the bright rings observed in some transition discs \citep[e.g.][]{2016A&A...596A..81P}, and also triggering streaming instabilities \citep[e.g.][]{2024A&A...686A..78S}. However, in numerical models these structures can be transient, especially if a radially-varying $\alpha(R)$ is combined with a smooth initial surface density profile. These structures typically evolve on the viscous time-scale $t_{\nu}$\,\footnote{Here, the viscous time-scale is typically $t_\nu= R^2/\nu \sim $Myr.}, and over time become much less effective dust traps. To avoid such numerical transients we allow the gas disc to relax to a self-consistent profile before introducing dust. We therefore let the gas-only disc evolve for $6~\mathrm{Myr}$, and the relaxed gas surface density is then taken as our initial condition in \textsc{DustPy} to co-evolve with the dust\footnote{The $6$-Myr gas evolution does not change the initial radial extent of gas and dust notably as a moderate turbulence $\alpha_\mathrm{SS}=10^{-3}$ is assumed for the majority of models. The size distribution of dust applied to the relaxed gas solution does not differ much from that to a self-similar solution.}.

\begin{figure}
	\centering
    \includegraphics[width=0.49\textwidth]{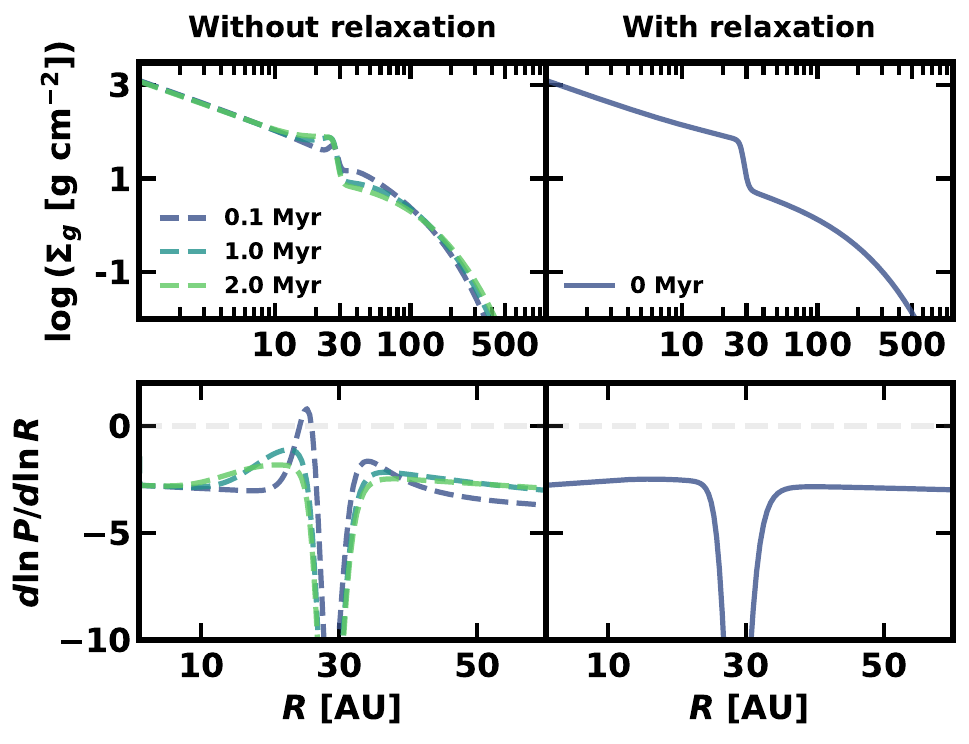}
    \caption{Evolution of the gas surface density and the pressure gradient ($d\ln P/d\ln R$) when a dead zone model is applied to a self-similar initial gas disc. The left panels show how the two quantities evolve without relaxation; the right panels illustrate the initial gas surface density and its pressure gradient adopted into our \textsc{DustPy} models, mimicking a disc where the dead zone develops gradually.}
    \label{fig:relax}
\end{figure}

The relaxation process is shown in Fig. \ref{fig:relax}. When the radius-dependent $\alpha_\mathrm{SS}(R)$ is initially inserted to the self-similar gas surface density, gas accumulates at the outer edge of the dead zone. This leads to a positive pressure gradient, which can effectively trap dust. The dust trap collapses gradually \citep{2016A&A...596A..81P} and remains as a boost to the pressure gradient, which can cause the `traffic jam' effect, for another a few million years. The relaxed gas surface density and its corresponding pressure gradient are shown in the right-hand side panels of Fig. \ref{fig:relax}. This gas surface density resembles those in \citet{2022A&A...658A..97D, 2023A&A...674A.190D}, where the gas profiles are computed self-consistently.

\begin{figure}
	\centering
    \includegraphics[width=0.49\textwidth]{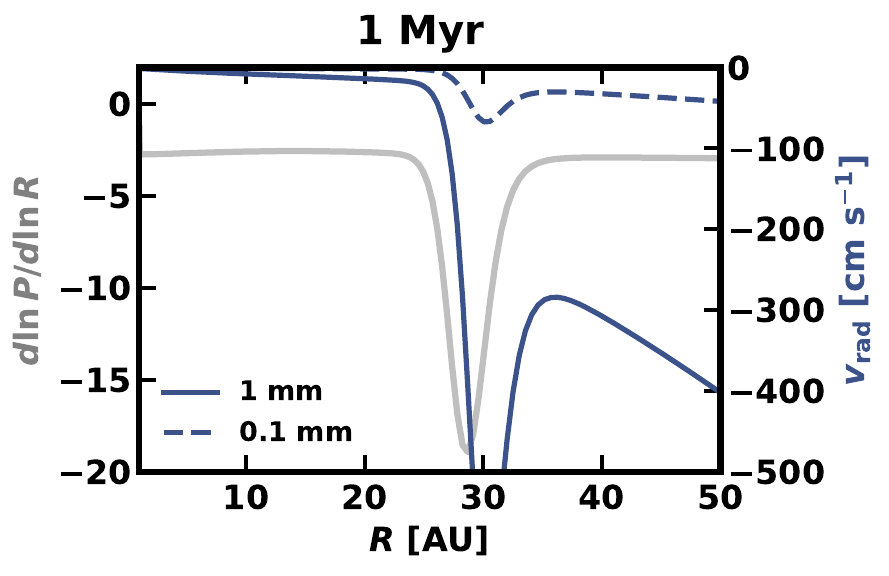}
    \caption{The pressure gradient (axis on the left-hand side in grey) and the radial velocity (axis on the right-hand side in dark blue) for mm (solid lines) and sub-mm (dashed lines) dust around the transition radius ($R_t=30~\mathrm{au}$). With the sharp transition in $\alpha(R)$, the radial velocity of 1\,mm dust exceeds $5\,\mathrm{m\,s^{-1}}$ at the outer edge of the dead zone. Data shown in this figure is from Model 2 (see Table \ref{tb:models1}).}
    \label{fig:relax_vrad}
\end{figure}

Although the transient dust trap dissipates, the sharp change in the pressure gradient at the dead zone edge can last most of disc lifetime. This can cause $\lesssim$mm-size dust to drift inwards rapidly (see Fig. \ref{fig:relax_vrad}, and also Eq. \ref{eq:dust_vr} and Fig. \ref{fig:relax}). Locally this radial drift can exceed the turbulent motion, and become the dominant dust velocity.
If the dust is fragile (i.e., $v_\mathrm{frag}\sim 1\,\mathrm{m\,s^{-1}}$), the disc thus falls into the drift-fragmentation regime (see blue lines in upper panels of Fig. \ref{fig:dust_distrib_fid}). We refer to this process as `hitchhiking'. We note that although the `hitchhiking' effect arises from dead zones in this case, any mechanisms that cause a sharp decrease in the gas surface density from the inner to the outer disc will have a similar effect.

We show in the following sections that overcoming the artificial pressure bumps can give rise to a radially compact protoplanetary disc, along with fragile dust (Section \ref{sec:setup}) and moderate turbulence expected in the MRI-active regions. We present our numerical models and analysis on them in Figs. \ref{fig:dust_distrib_fid}-\ref{fig:MR_M6-11} and show the observational appearance of these models in Fig. \ref{fig:mock_obs}.

\begin{table*}
\caption{Input parameters and resulting dust disc properties at $\lambda=1.3\,\mathrm{mm}$ for models without pressure bumps, evaluated at $t=1$ \& $3\,\mathrm{Myr}$ and computed using \textsc{DSHARP} opacities \citep{2018ApJ...869L..45B}.}\label{tb:models1}
\begin{tabular}{cccccccccccc}
\hline
\hline
(1) & (2) & (3)  & (4)  & (5)  & (6)  & (7)  &(8) &(9) &(10)& (11) & (12)\\
{Model} & {$\alpha_\mathrm{DZ}$} & {$\alpha_\mathrm{MRI}$} &  $R_t$  &  $M_\mathrm{d, tot}^\mathrm{1Myr}$&$F_\mathrm{1.3mm}^\mathrm{1Myr}$ & $R_\mathrm{dust,68\%}^\mathrm{1Myr} $& $R_\mathrm{dust,90\%}^\mathrm{1Myr}$&  $M_\mathrm{d, tot}^\mathrm{3Myr}$&  $F_\mathrm{1.3mm}^\mathrm{3Myr}$ & $R_\mathrm{dust,68\%}^\mathrm{3Myr} $& $R_\mathrm{dust,90\%}^\mathrm{3Myr}$\\
  & &  & [au] & [$M_\oplus$] & [mJy] & [au] & [au] & [$M_\oplus$]& [mJy] & [au] & [au]  \\ 
\hline
1 & $10^{-4}$ & $10^{-2}$ & 30 & 8.97 & 9.88 & 13.37 & 23.77 & 3.74 & 2.82 & 18.88 & 23.77 \\
2 & $10^{-4}$ & $10^{-3}$ & 30 & 61.76 & 19.71 & 16.83 & 26.67 & 44.29 & 15.15 & 16.83 & 23.77 \\
3 & $10^{-4}$ & $5\times 10^{-4}$ & 30 & 76.36 & 30.91 & 18.88 & 29.93 & 43.02 & 27.38 & 16.83 & 26.67 \\
4 & $10^{-4}$ & $10^{-3}$ (S) & 30 & 121.09 & 97.71 & 26.67 & 37.68 & 52.03 & 59.79 & 21.19 & 29.93 \\
5 & $10^{-4}$ & $10^{-3}$ & 20 & 73.97 & 15.22 & 13.37 & 18.88 & 55.88 & 13.71 & 13.37 & 18.88 \\
6 & $10^{-4}$ & $10^{-3}$ & 50 & 67.25 & 35.95 & 16.83 & 33.58 & 28.82 & 16.11 & 23.77 & 37.68 \\
\hline
\end{tabular}
\end{table*}

\subsection{Dust evolution}\label{sec:result_evo}
Models 1--6 study how dead zone models affect disc evolution by varying $\alpha_\mathrm{SS,MRI}$ (Models 1-3), the transition radius (Models 2, 5 and 6), and the sharpness of the transition (Models 2 and 4), with a low fragmentation velocity $v_\mathrm{frag}=1~\mathrm{m~s^{-1}}$, in discs with no additional pressure bumps. In each case we examine the dust properties, including the total dust mass ($M_\mathrm{d,tot}$), the observed continuum fluxes at $1.3~\mathrm{mm}$ ($F_\mathrm{1.3mm}$), and the (dust) radii enclosing $68\%$ ($R_{d,\mathrm{68\%}}$) and $90\%$ ($R_{d,\mathrm{90\%}}$) of the total continuum fluxes for these models. The model parameters and these results are presented in Table \ref{tb:models1}.

\subsubsection{Formation of compact discs}\label{sec:result_formation}

\begin{figure*}
    \centering
    \includegraphics[width=1.0\textwidth]{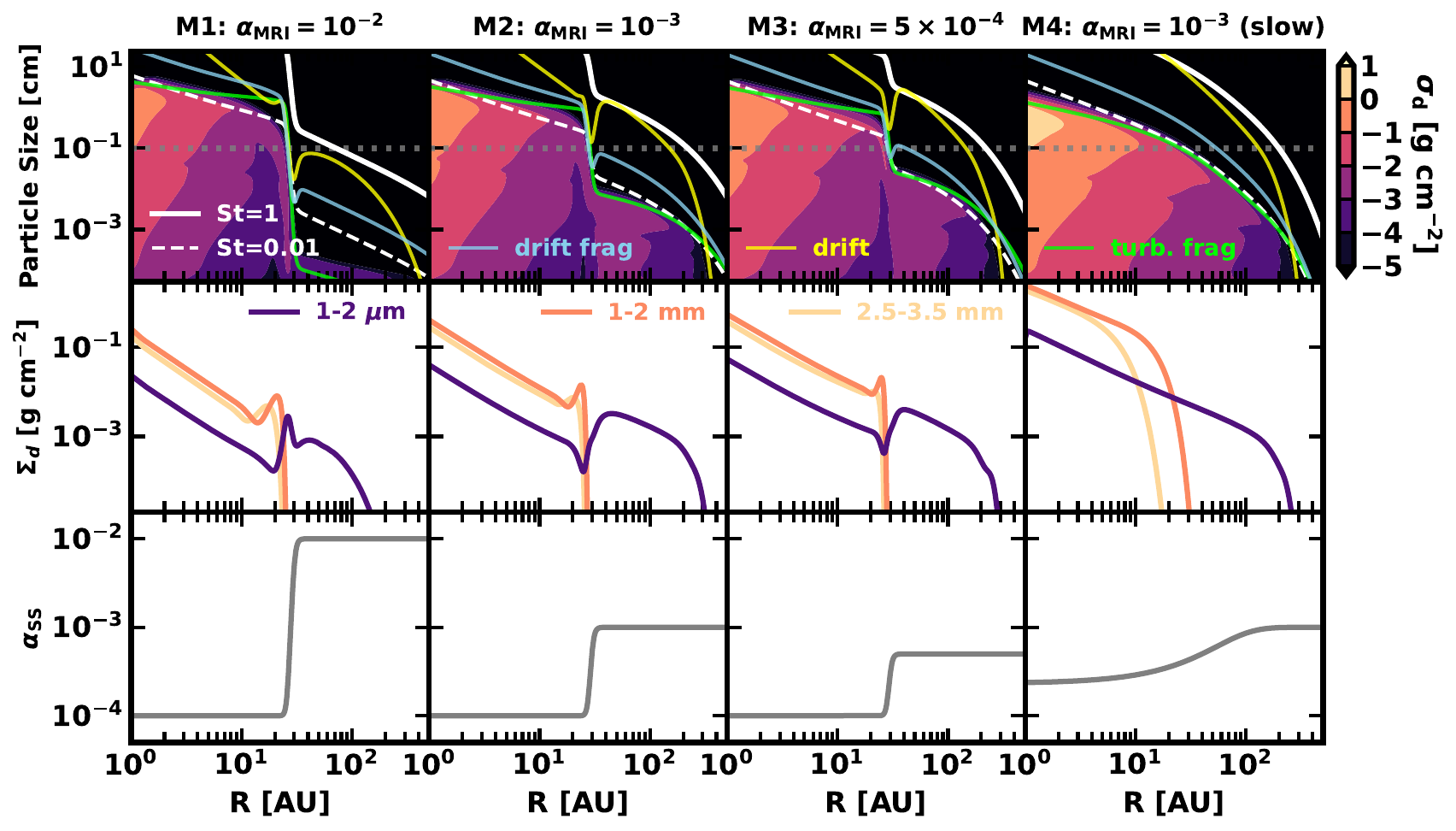}
    \caption{The dust distributions and transition profiles for Models 1-4 (from left to right) at $1~\mathrm{Myr}$. Upper panels: dust density distributions as a function of radius. Middle panels: integrated surface densities of $1$-$2~\mathrm{\mu m}$ (purple), $1$-$2~\mathrm{mm}$ (orange) and $2.5$-$3.5~\mathrm{mm}$ (yellow) dust. Lower panels: the transition of $\alpha_\mathrm{SS}$ from dead zones ($r<30~\mathrm{au}$) to MRI-active zones ($r>30~\mathrm{au}$). In the upper panels the green and blue solid lines show the maximum particle size set by turbulence-dominated and drift-dominated fragmentation, respectively. The yellow solid lines are the maximum particle size limited by radial drift. The white solid and dashed lines signify the particle sizes with $\mathrm{St}=1$ and $10^{-2}$, respectively. The dotted horizontal grey line denotes dust with a size of $1~\mathrm{mm}$.}
    \label{fig:dust_distrib_fid}
\end{figure*}

Fig. \ref{fig:dust_distrib_fid} shows the dust distributions for Models 1--4. Even moderate turbulence in the outer disc prevents dust growth to millimetre sizes, so almost no mm-size dust is seen in the MRI-active region in all four models. Dust fragility therefore provides an alternative way to form compact discs, which is distinct from the previously proposed mechanisms of drift-dominated \citep[e.g.][]{2019A&A...626L...2F} and born-small discs.

Models 1--3 assume various $\alpha_\mathrm{SS,MRI}$, ranging from $10^{-2}$ to $5\times 10^{-4}$. These values are within the reasonable range of $\alpha_\mathrm{SS}$ in outer discs inferred from ALMA observations \citep{2023NewAR..9601674R}. When most of the disc is fragmentation-limited, no mm-dust is formed in the MRI-active region. The difficulty of growing mm-dust in turbulent discs with $\alpha_\mathrm{SS}=10^{-2}$ and $10^{-3}$ has been reported in 
\citet[][see their Models 1 and 7 in Fig. 2]{2021A&A...645A..70P}. For $\alpha_\mathrm{SS}=10^{-2}$, forming mm dust is difficult even for an extreme fragmentation velocity  \citep[$v_\mathrm{frag}=10\,\mathrm{m\,s^{-1}}$, e.g.][]{2009A&A...503L...5B}. However, our models struggle to form mm-dust even for $\alpha_\mathrm{SS}= 5\times 10^{-4}$ in the outer disc (Model 3), which is lower than the majority of $\alpha_\mathrm{SS}$ derived from current ALMA observations. We discuss this further in Section \ref{sec:disc_var_alpha}.

Some recent non-ideal MHD simulations show that the transition of $\alpha_\mathrm{SS}$ from the MRI-quenched to MRI-active regions can be quite smooth, with no mass accumulated at the dead zone outer edge \citep[e.g.][]{2016ApJ...818..152B, 2017A&A...600A..75B}. We therefore also consider a smooth transition profile \citep[Model 4, which resembles the lower left panel of Figure 2 in][]{2016ApJ...818..152B}. For consistency with other models in Fig. \ref{fig:dust_distrib_fid}, we move the transition radius to $30~\mathrm{au}$ in this case.

The dust density distribution of a disc with a slow transition (Model 4) is not significantly different from models that adopt a globally constant $\alpha_\mathrm{SS}$ (see the model without bumps in \citet{2022A&A...668A.104S}, where $v_\mathrm{frag}=10~\mathrm{m~s^{-1}}$ and models 1/2/7/9 in \citet{2021A&A...645A..70P}, where $v_\mathrm{frag}=1~\mathrm{m~s^{-1}}$), except for the lack of mm dust inside the characteristic radius $R_c$. This is the result of gradually increasing turbulence to a sufficiently high value at $R<R_c$ for Model 4. 

The model with a smooth transition profile (Model 4) differs from models with sharp transition profiles (Model 1-3) in two aspects (see Fig. \ref{fig:dust_distrib_fid}): i) the edges of 1-2 mm and 2.5-3.5 mm dust discs differ in Model 4 and are similar in Models 1-3 (around the transition radius $R_t$), as the edges depend on where turbulence becomes so high that dust of corresponding sizes cannot survive; ii) the slightly increased surface densities of mm-dust at the outer edge, caused by the slowing down of the radial velocity following `hitchhiking', is found in Models 1-3, but is not seen in Model 4. This is because the slow transition in Model 4 does not lead to a sharp drop in gas surface densities. These two signatures can be employed to distinguish discs with a slow transition from those with a sharp one in multi-wavelength observations.

The former appear smaller for the radial extent of larger particles, while the latter remain almost unchanged (see the middle panels of Fig. \ref{fig:dust_distrib_fid}). This trend is also seen in $R_{d,\mathrm{90\%}}$ measured from synthetic observations at $\lambda=1.3$ and $3~\mathrm{mm}$, and is prominent when $\alpha_\mathrm{SS, MRI}\gtrsim 10^{-3}$. The independence of observed disc sizes on the wavelengths coincides with recent multi-wavelength observations \citep[e.g.][]{2021MNRAS.506.2804T, 2022ApJ...930...56U, 2023A&A...673A..77R}. A slow transition also results in a less sharp dust outer edge in the radial intensity of the mock observations (see Fig. \ref{fig:radial_intensity_all}). The increased dust surface density at the disc outer edge in Models 1-3 is reflected as a low-contrast ring (noise-free) or a `shoulder' (noise-included) in the radial intensity profiles of synthetic observations (Fig. \ref{fig:radial_intensity_all}).

It is worth noting that lack of mm-dust in the MRI-active region does not mean the disc has no emission from that region at mm wavelengths. However, the weak emission from smaller grains is comparable to the typical noise level in the synthetic ALMA observations (which assume an integration time of 30 mins and a beam size of $0.^{\prime \prime}05$), and is therefore unlikely to be detectable.

\subsubsection{Size of the compact disc}\label{sec:result_size}

The transition radius of the dead zone depends on a number of physical parameters, including the ionization and the strength and geometry of the magnetic field, which vary between discs, and as individual discs evolve \citep{2023A&A...674A.190D}. In this section we study the sizes of compact discs formed in the scenario proposed above. We simply fix the transition radius to $R_t=20$ (Model 5), $30$ (Model 2) and $50~\mathrm{au}$ (Model 6), following results from numerical calculations \citep[e.g.][]{2000ApJ...543..486S, 2013ApJ...765..114D, 2022A&A...658A..97D}, and keep other parameters the same as in Model 2. 

\begin{figure*}
    \centering
    \includegraphics[width=0.78\textwidth]{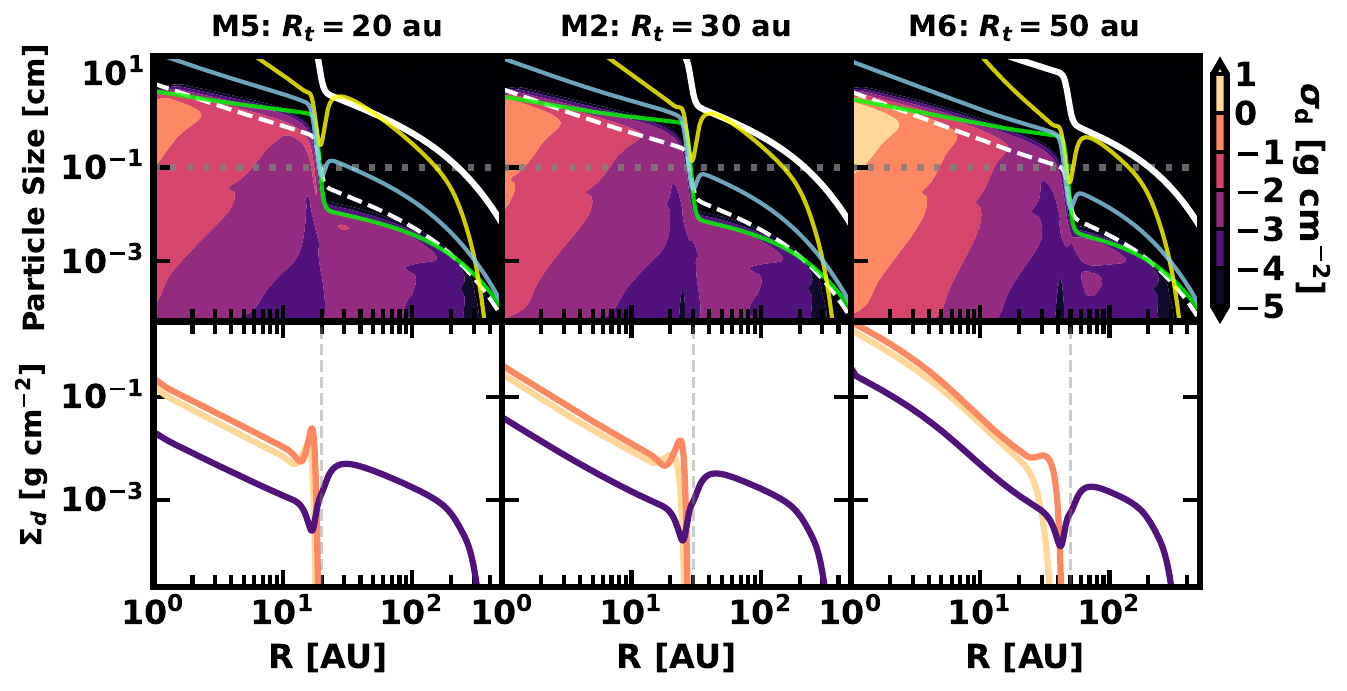}
    \caption{Dust distributions plotted as in Fig. \ref{fig:dust_distrib_fid} but for Models 5 (left, $R_t=20~\mathrm{au}$), 2 (middle, $R_t=30~\mathrm{au}$) and 6 (right, $R_t=50~\mathrm{au}$) at $1~\mathrm{Myr}$. Each of them adopts a dead zone model with different transition radius $R_t$. The vertical dashed grey lines in the lower panels indicate the outer edge of dead zones.}
    \label{fig:dust_distrib_fid_rt}
\end{figure*}

\begin{figure*}
    \centering
    \includegraphics[width=0.95\linewidth]{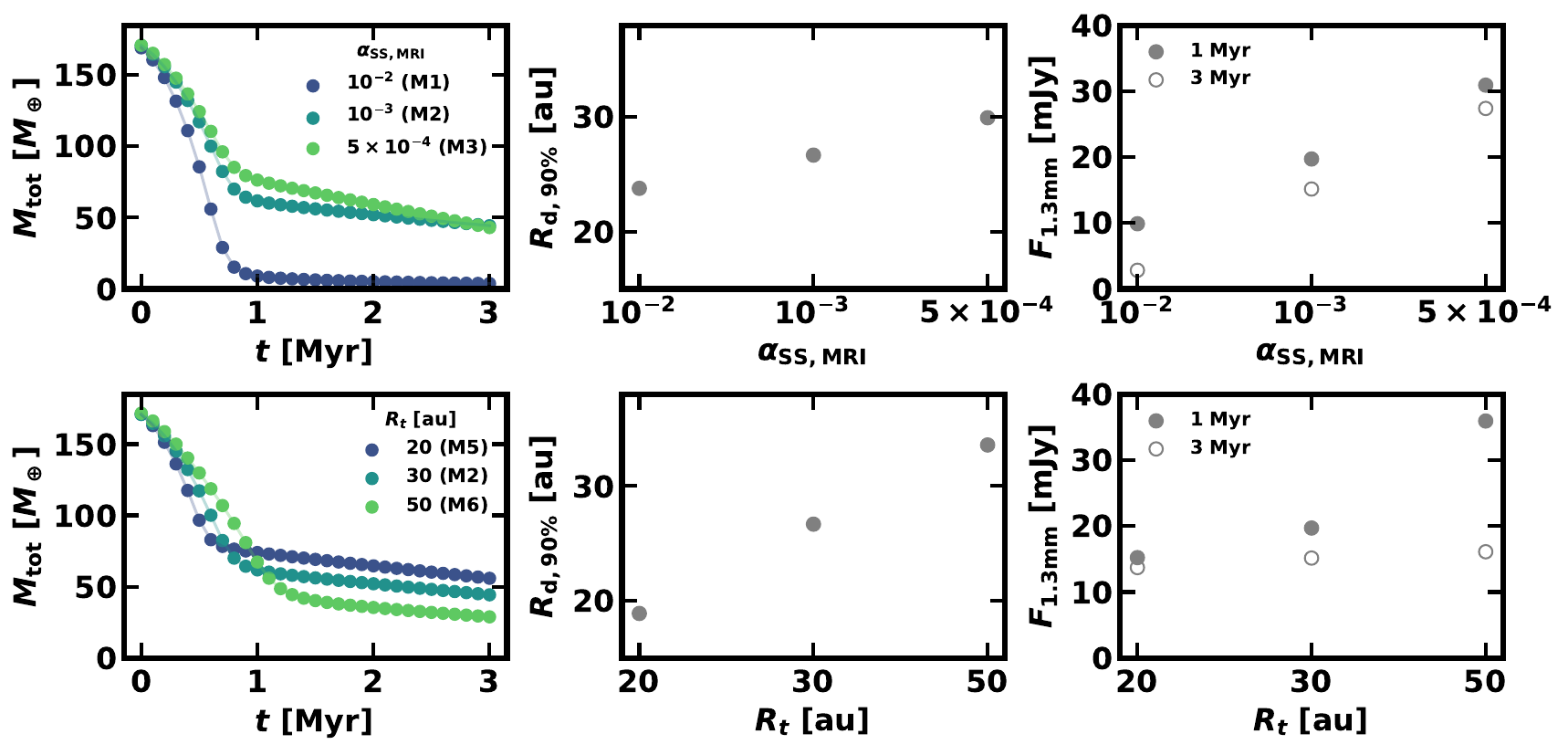}
    \caption{Evolution of the dust total mass (left panels); the observed disc radii $R_\mathrm{d,{90\%}}$ at 1 Myr (middle panels); and the measured continuum fluxes $F_\mathrm{1.3mm}$ at 1 and 3 Myr (right panels) for Models 1-3, 5 and 6. The upper panels show the results from models with different values of $\alpha_\mathrm{SS, MRI}$ (Models 1-3); the lower panels are for models with different $R_t$ (Models 2, 5 and 6).}
    \label{fig:MR_M1-5}
\end{figure*}

A comparison between these models is shown in Fig. \ref{fig:dust_distrib_fid_rt}. The evolution of the total dust mass, and various observables, are shown in Fig. \ref{fig:MR_M1-5}. Fig. \ref{fig:dust_distrib_fid_rt} shows that the radial extent of the mm-dust is limited by the transition radius, indicating that discs with larger dead zones have larger sizes. The same effect is seen in the observed disc sizes $R_{d,\mathrm{ 90\%}}$, in the lower middle panel of Fig. \ref{fig:MR_M1-5}. In contrast, $R_{d,\mathrm{ 90\%}}$ is only weakly dependent on $\alpha_\mathrm{SS,MRI}$ (see the upper middle panel of Fig. \ref{fig:MR_M1-5}). The disc size increases only slightly when $\alpha_\mathrm{SS,MRI}$ is reduced by two orders of magnitude, as dust is better retained for discs with low $\alpha_\mathrm{SS,MRI}$. 

$R_{d,\mathrm{90\%}}$, as a more accurate indicator of disc radial extent than $R_{d,\mathrm{68\%}}$, is almost constant for Model 1-3 and Model 5-6 during the 3-Myr evolution. The implications of the unchanged disc radius over time are discussed in Section \ref{sec:disc_SLR}. $R_{d, \mathrm{68\%}}$ instead traces an increasing dust size for Model 1, which has a large $\alpha_\mathrm{SS,MRI}=10^{-2}$, consistent with the prediction in \citet{2019MNRAS.486.4829R}.

On the other hand, the observed flux $F_\mathrm{1.3mm}$ depends more strongly on $\alpha_\mathrm{SS,MRI}$ than $R_t$. Discs with lower $\alpha_\mathrm{SS,MRI}$ have good retention of dust (the upper left panel of Fig. \ref{fig:MR_M1-5}), maintaining a consistently higher flux over the entire evolution ($3~\mathrm{Myr}$). Although discs with larger $R_t$ but the same $\alpha_\mathrm{SS}$ have initially higher fluxes, their fluxes drop significantly after a few million years, due to the decrease in the total dust mass and the optical depth (lower left panels of Fig. \ref{fig:MR_M1-5}). At $3\,\mathrm{Myr}$, discs with different $R_t$ but the same other parameters have essentially the same continuum flux. 

\subsection{Presence of planetary gaps in gas discs}\label{sec:result_bump}

Planetary mass objects can alter the gas surface density and create dust traps, slowing down the radial drift of large particles and enhancing dust concentration. In this section, we mimic the effect of planets on discs to investigate how they change the disc morphology at mm wavelengths. We insert Gaussian bumps into $\alpha_\mathrm{SS}$ in Model 2, using the prescription in Eq. \ref{eq:bump}. We set $R_\mathrm{gap}$ to be $6$, $15$, $30$, $45$ and $80~\mathrm{au}$ and the amplitudes of these Gaussian bumps $A_\mathrm{gap}$ increase with radius to maintain a positive pressure gradient at similar levels. The parameters of these models are listed in Table \ref{tb:models+bumps}.

\begin{table*}
\caption{Input parameters and resulting dust disc properties at $\lambda=1.3\,\mathrm{mm}$ for models with pressure bumps, evaluated at $t=1$ \& $3\,\mathrm{Myr}$ and computed using \textsc{DSHARP} opacities \citep{2018ApJ...869L..45B}.}\label{tb:models+bumps}
\begin{tabular}{cccccccccccc}

\hline
\hline
(1) & (2) & (3)  & (4)  & (5)  & (6)  & (7)  &(8) &(9) &(10)& (11) &(12)\\
{Model} & {$R_\mathrm{gap}$} & {$A_\mathrm{gap}$} & Planetary mass&$M_\mathrm{d, tot}^\mathrm{1Myr}$&$F_\mathrm{1.3mm}^\mathrm{1Myr}$ & $R_\mathrm{dust,68\%}^\mathrm{1Myr} $& $R_\mathrm{dust,90\%}^\mathrm{1Myr}$&  $M_\mathrm{d, tot}^\mathrm{3Myr}$&  $F_\mathrm{1.3mm}^\mathrm{3Myr}$ & $R_\mathrm{dust,68\%}^\mathrm{3Myr} $& $R_\mathrm{dust,90\%}^\mathrm{3Myr}$\\
  & [au]  & & &[$M_\oplus$]& [mJy] & [au] & [au] & [$M_\oplus$]& [mJy] & [au] & [au]  \\ 
\hline
7 & 6 & 0.7 & $3.12~M_\oplus$ & 129.72 & 23.06 & 13.37 & 23.77 & 74.37 & 20.02 & 11.91 & 23.77 \\ 
8 & 15 & 1.2& $6.94~M_\oplus$ & 116.26 & 19.77 & 21.19 & 26.67 & 113.93 & 19.65 & 21.19 & 26.67 \\
9 & 30 & 1.5 & $0.11~M_\mathrm{Jup}$ & 63.80 & 15.41 & 15.00 & 23.77 & 49.42 & 14.06 & 16.83 & 23.77 \\
10 & 45 & 2 & $0.21~M_\mathrm{Jup}$ & 60.37 & 13.49 & 13.37 & 23.77 & 49.93 & 10.17 & 16.83 & 26.67 \\ 
11 &80 & 3 & $0.33~M_\mathrm{Jup}$ & 61.48 & 26.53 & 18.88 & 29.93 & 43.01 & 8.08 & 18.88 & 26.67 \\
\hline
\end{tabular}
\end{table*}

\begin{figure*}
    \centering
    \includegraphics[width=1.0\textwidth]{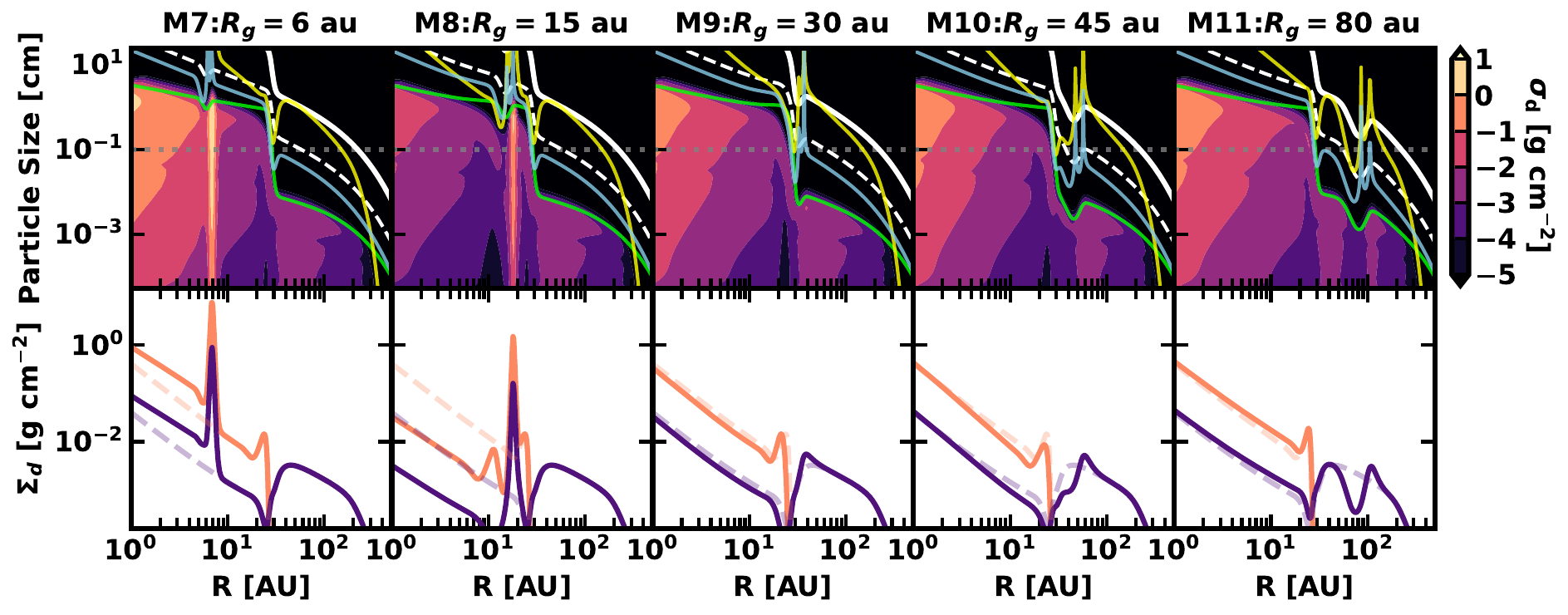}
    \caption{Dust distributions plotted as in Fig. \ref{fig:dust_distrib_fid} for Models 7-11, at $t=1\,\mathrm{Myr}$. These are identical to Model 2, but with additional traps at $6$, $15$, $30$, $45$ and $80~\mathrm{au}$ (from left to right). Upper panels: dust density distributions as a function of radius. Lower panels: integrated surface densities of $1$-$2~\mathrm{\mu m}$ (purple) and $1$-$2~\mathrm{mm}$ (orange) dust. The overlaid transparent dashed lines are from Model 2 (with no traps). The coloured lines denote the same physical quantities as those in Fig. \ref{fig:dust_distrib_fid}.}
    \label{fig:dust_distri_bumps1}
\end{figure*}

Fig. \ref{fig:dust_distri_bumps1} shows the dust distributions of Models 7--11. The inclusion of dust traps inside the dead zone (Model 7 and 8) accumulates mm- and $\mathrm{\mu m}$- dust, and creates a ring-like feature in the dust surface density. This contrasts with models including dust traps in the MRI-active region (Models 9-11), which do not differ substantially from discs without traps (Model 2). Traps in the MRI-active region only enhance the accumulation of micron-dust, and have little effects on mm-dust, which is not formed without traps. The enhanced micron-dust creates a ring feature, but its emission at mm wavelengths is comparable to the noise level in ALMA observations (see Figure \ref{fig:radial_intensity_all}). The outer ring therefore remains nearly unseen in synthetic observations. The only visible ring that is associated with the added pressure bump appears in Model 8 (see Fig. \ref{fig:mock_obs}). 

The ring in Model 7 (at $\gtrsim 6~\mathrm{au}$), which is prominent in the numerical model but invisible in the radiative transfer model and synthetic observations, is the result of the optically thick dust. Particles with larger Stokes numbers can be formed around the pressure bump in the inner discs. They collide with higher relative velocities and are more likely to fragment to small grains. These small grains are not trapped as effectively as their larger counterparts, and drift inwards to replenish the inner disc, leading to a leaky trap, which has also been found in other models \citep[e.g.][]{2023A&A...670L...5S} and probably captured in observations \citep[e.g.][]{2017ApJ...839...99P, 2019ApJ...878...16P, 2024A&A...682A..55M}. The inward drifting small grains enhance the dust surface density within $6~\mathrm{au}$, and result in a high optical depth. 

\begin{figure}
    \centering
    \includegraphics[width=0.9\linewidth]{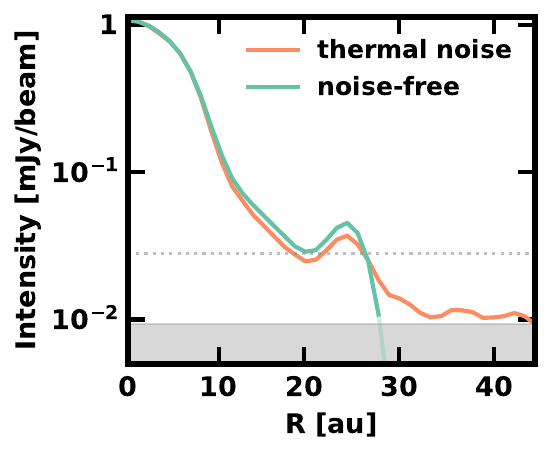}
    \caption{The azimuthally-averaged radial intensity profile extracted from the synthetic observation of Model 7. The beam size is $0.\arcsec 02$ and the integration time is $30$ mins. The grey shade indicates the root-mean-square noise $\sigma$ measured from noisy images, and the grey dotted line indicates the level of $3\sigma$.}
    \label{fig:radial_M7}
\end{figure}

The disc beyond $\gtrsim 6~\mathrm{au}$ is optically thin and has lower emission than the optically thick part. When the resolution is moderate ($0.\arcsec05$), the decreased emission blends with the weak `hitchhiking' ring, showing as a `shoulder' feature (Fig. \ref{fig:radial_intensity_all}). When the resolution is high ($0.\arcsec02$), the two features are separately resolved as in Fig. \ref{fig:radial_M7}. The change in the slope of the radial intensity profile marks the transition from the optically thick to thin regions, and a more evident ring from the `hitchhiking' effect is shown at the disc outer edge. The radial intensity profile of Model 7 is strikingly similar to that recently observed in Sz 66 \citep[][their Fig. 13, where the synthesized beam is $0.\arcsec026\times0.\arcsec018$]{2024A&A...682A..55M}. Similar slope changes and a weak ring at the outermost radius have also been observed in MP Mus, a seemingly featureless disc with $R_\mathrm{d,90\%}=45~\mathrm{au}$ \citep{2023A&A...673A..77R}.

\begin{figure*}
    \centering
    \includegraphics[width=0.95\linewidth]{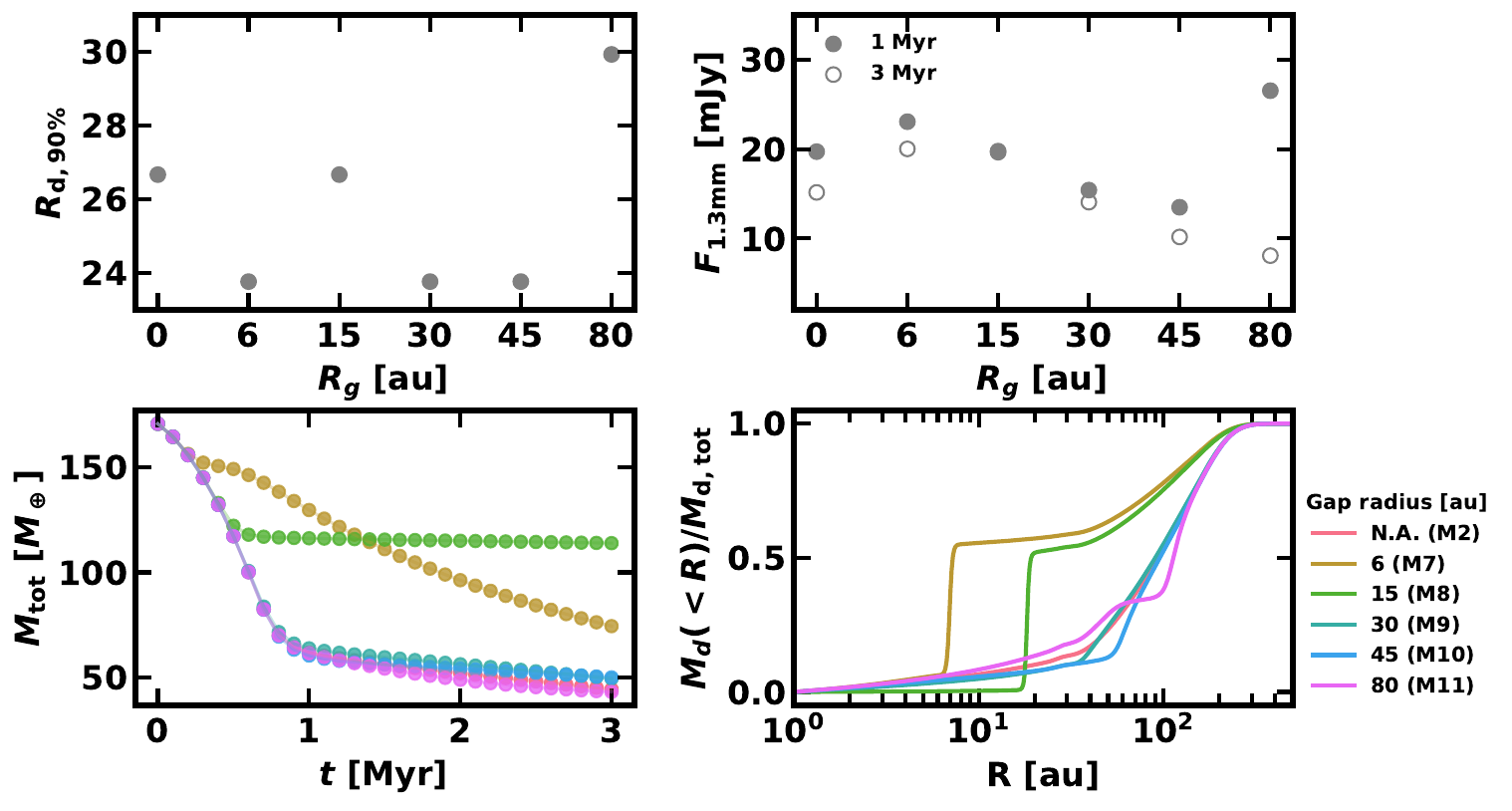}
    \caption{Upper left: the measured dust disc size $R_{d,\mathrm{90\%}}$ at $\lambda=1.3\,\mathrm{mm}$ for models without ($R_g=0~\mathrm{au}$, Model 2) and with ($R_g> 0~\mathrm{au}$, Models 7-11) traps at $t=1\,\mathrm{Myr}$. Upper right: The measured continuum flux at $1.3~\mathrm{mm}$ for models with and without traps, plotted at 1 Myr (filled grey circles) and 3 Myr (open grey circles). Both markers for Model 8 ($R_g \sim 15~\mathrm{au}$) overlap, as its continuum flux remains unchanged during this period. Lower left: The evolution of the total dust mass for models without (pink) and with traps around $6$ (gold), $15$ (green), $30$ (cyan), $45$ (blue), $80~\mathrm{au}$ (magenta). Lower right: The normalized cumulative mass distributions for Models 2, 7-11 (the colour scheme is the same as that in the lower left panel).} 
    \label{fig:MR_M6-11}
\end{figure*}

\begin{figure*}
    \centering
    \includegraphics[width=0.99\textwidth]{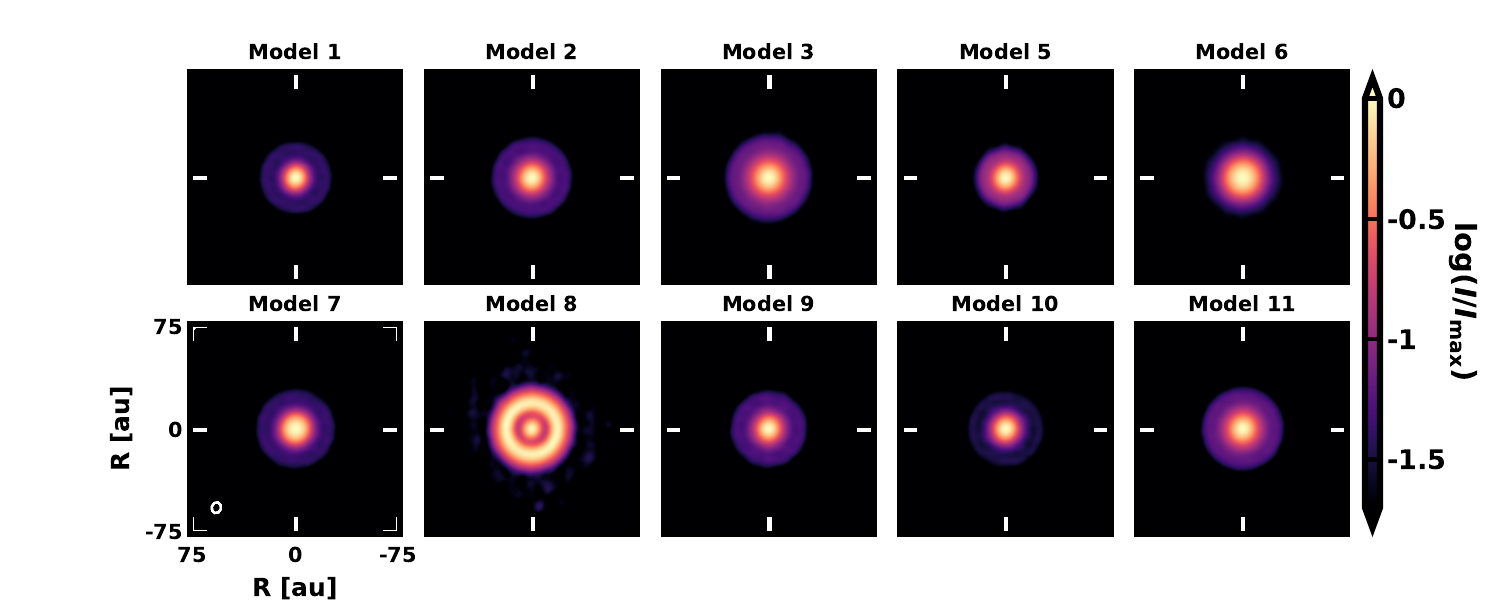}
    \caption{ALMA synthetic observations at $\lambda = 1.3~\mathrm{mm}$ for models in Table \ref{tb:models1} and \ref{tb:models+bumps} at $t=1\,\mathrm{Myr}$ (integration time: 30 min, resolution $\sim 0.^{\prime\prime}05$). The top and bottom panels are for models without and with traps, respectively. The faint rings at the outer edge are caused by the `hitchhiking' effect introduced in Section \ref{sec:result_DZ_revis}, and cannot be detected once noise is considered. The beam size is shown as a white ellipse in the bottom row of the first column. (Model 4, which assumes a smooth transition, is not shown here.)}
    \label{fig:mock_obs}
\end{figure*}

Regardless of the disc morphology, the imposition of dust traps in all the models shown here barely affects the radial extent of the mm- and micron-dust, as shown in the lower panels of Fig. \ref{fig:dust_distri_bumps1}. However, traps present at $80~\mathrm{au}$ can slightly enlarge the observed disc radius ($R_{d,\mathrm{68\%}}$ and $R_{d,\mathrm{90\%}}$) as slightly more dust is distributed at larger radii (the upper left panel of Fig. \ref{fig:MR_M6-11}). 

Traps in the MRI-active regions also have little or no effect on dust mass retention. In the lower left panel of Fig. \ref{fig:MR_M6-11}, models with traps in the MRI-active region have almost the same total mass as the model with no trap at any time during the 3-Myr evolution. By contrast, Models 7 and 8, which have traps within the MRI-suppressed regions, can retain more than $40$ per cent of the initial mass after $3\,\mathrm{Myr}$, and accumulate $\sim 50$ per cent of the total dust mass around the trap at $1\,\mathrm{Myr}$ (the lower right panel of Fig. \ref{fig:MR_M6-11}).

Models which fail to retain sufficient dust also fail to retain high continuum fluxes. Models 7 and 8, with dust traps in the dead zone, have higher fluxes during the $3$-$\mathrm{Myr}$ evolution than other models (the upper right panel of Fig. \ref{fig:MR_M6-11}). When the trap is close to the central star it receives more dust supply from further out, making the disc brighter. This aligns with findings in \citet{2022A&A...668A.104S}. Models 9-11 have dust traps in the MRI-active outer disc, where the dust supply is small and lacks large particles that can be efficiently trapped. This yields an observable flux, which is heavily dependent on the optical depth. This explains the exceptionally high flux $F_\mathrm{1.3mm}$ in Model 11 at $1~\mathrm{Myr}$, and its significant drop over $3~\mathrm{Myr}$. 
 
\section{Discussion}\label{sec:discussion}
\subsection{Size-luminosity relation}\label{sec:disc_SLR}
The observed size-luminosity relation has been investigated in a handful of studies \citep[e.g.][]{2017A&A...606A..88T} and the relation  $R_\mathrm{eff}\propto L_\mathrm{mm}^{0.5}$ is now well established \citep[e.g.][]{2017ApJ...845...44T,  2018ApJ...865..157A}. Though the observed size-luminosity relation has been proposed as a natural result of drift-dominated discs \citep{2019MNRAS.486.4829R}, it is also commonly found in radially extended discs with substructures \citep{2022A&A...661A..66Z, 2024A&A...688A..81D} and in compact discs, which may include (marginally) resolved substructures \citep[e.g.][]{2019ApJ...882...49L, 2021A&A...645A.139K, 2024ApJ...966...59S}. Circumstellar discs in binary systems from Taurus \citep{2019A&A...628A..95M} are also marginally consistent with this relation \citep{2021MNRAS.507.2531Z}\footnote{We note that analysis also by \citet{2021MNRAS.507.2531Z} suggests a shallower relation in $\rho$-Ophiuchus \citep{2017ApJ...851...83C}. They attributed this to the lower-quality observations or the intrinsic property of this region.}. 

\begin{figure}
    \centering
    \includegraphics[width=1\linewidth]{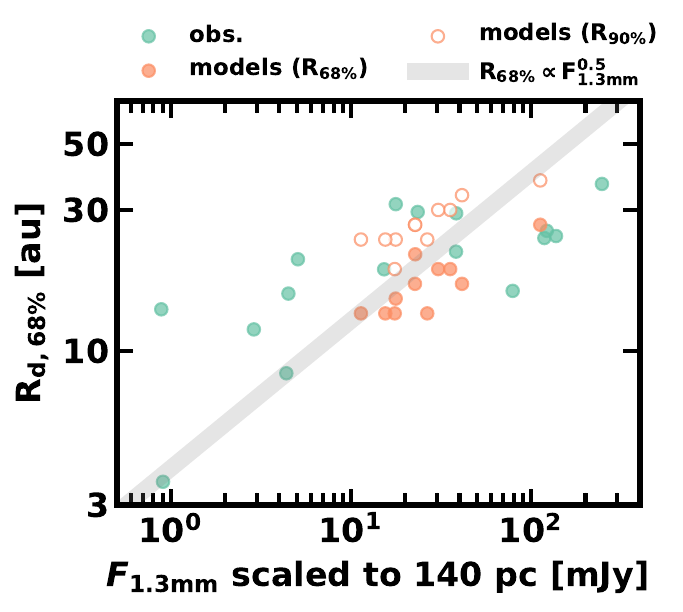}
    \caption{The size-luminosity relation for models from Section \ref{sec:result} (orange open circles are for $R_\mathrm{d. 90\%}$ and orange filled circles are for $R_\mathrm{d. 68\%}$) and for compact discs ($\lesssim 50~\mathrm{au}$) observed in ALMA Band 6 \citep{2019ApJ...882...49L, 2024ApJ...966...59S} (green dots). The grey line shows the proportionality between disc sizes and fluxes $R_\mathrm{68\%}\propto F_\mathrm{1.3mm}^{0.5}$ \citep[e.g.][]{2017ApJ...845...44T, 2018ApJ...865..157A} for the purpose of eye guidance only. The two orange dots with $F_\mathrm{1.3mm}>100~\mathrm{mJy}$ are from Model 4, which assumes a smooth transition from MRI-inactive to MRI-active regions.}
    \label{fig:SLR}
\end{figure}

In this section, we examine whether compact discs formed by the dead zone, which fall in the fragmentation-limited regime, also follow the same size-luminosity relation. We scale fluxes $F_\mathrm{1.3mm}$ to $140~\mathrm{pc}$ and measure dust sizes ($R_{d,\mathrm{68\%}}$ and $R_{d,\mathrm{90\%}}$) for the models discussed in Section \ref{sec:result} at $1~\mathrm{Myr}$, and plot them in Fig. \ref{fig:SLR}. The overlaid samples are observed compact discs ($<50~\mathrm{au}$) from \citet{2019ApJ...882...49L} and \citet{2024ApJ...966...59S}\footnote{Most smooth discs in \cite{2019ApJ...882...49L} are found to include substructures, such as rings, gaps and shoulders, by visibility modelling \citep{2023ApJ...952..108Z, 2024PASJ...76..437Y}. Discs in \citet{2024ApJ...966...59S}, except 2M0450 and CIDA12, are also found to have substructures, such as gaps, rings and inner cavities.}. Our models generally follow a similar trend to the observations, but are clustered due to the small parameter space explored here. This implies that compact discs formed by dead zones are consistent with the observed size-luminosity relation. However, given the small range explored by our models, we do not fit an explicit power-law relation to our results.

A more flattened size-luminosity relation has been found for discs strongly irradiated by nearby massive stars \citep{2018ApJ...860...77E, 2021ApJ...923..221O} and for discs from older star-forming regions, such as Upper Sco \citep{2020ApJ...895..126H}, though discs in Upper Sco may also experience mild external irradiation today (Anania et al, submitted) and have experienced stronger irradiation when they were more clustered historically. If the flatter relation can be at least partly attributed to disc evolution, the shallow slope may result from reduced continuum fluxes while disc sizes are maintained \citep{2021A&A...645A.139K}. This is consistent with the continuum fluxes and measured dust sizes of our models with sharp transitions (see Table \ref{tb:models1} and \ref{tb:models+bumps}, except Model 4). Note, however, that our models are only evolved to $3~\mathrm{Myr}$, which is younger than the estimated age of Upper Sco \citep[$5-11~\mathrm{Myr}$, ][]{2002AJ....124..404P, 2012ApJ...746..154P}. 

In addition to matching the observed size-luminosity relation, our models predict that compact discs formed by dead zones have a high gas-to-dust size ratio, which is usually thought to be characteristic of drift-dominated discs \citep[e.g.][]{2019A&A...626L...2F, 2020A&A...638A..38T}. While the dust size is constrained by the dead zone outer edge, which shrinks as the gas density drops and the opacity is reduced \citep[e.g.][]{2023A&A...674A.190D}, the outer gas disc continues spreading if $\alpha_\mathrm{SS,MRI}$ is not too low \citep{2024_disc_evo}\footnote{The effect of MHD winds is not considered here.}, leading to an increasing gas-to-dust size ratio over time. Unfortunately, quantitative analysis of this trend requires detailed knowledge of the turbulence in the outer disc, which is beyond the scope of our current models.

\subsection{Visible dust rings at small radii}\label{sec:disc_invisble}

\begin{figure}
    \centering
    \includegraphics[width=1.0\linewidth]{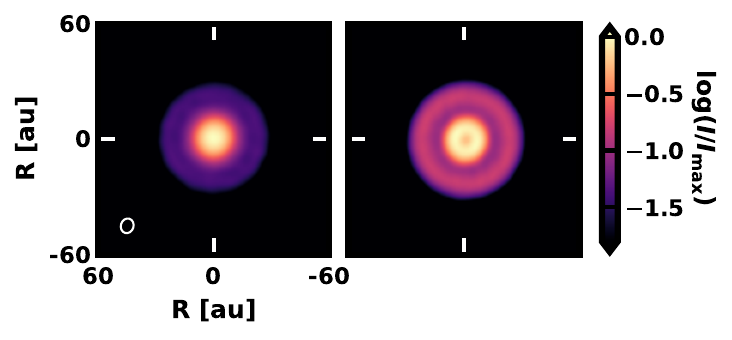}
    \caption{The synthetic observations at $\lambda = 1.3~\mathrm{mm}$ for Model 7 ($v_\mathrm{frag}=1~\mathrm{m~s^{-1}}$, left) and modified Model 7 (M7$^{\prime}$,$v_\mathrm{frag}=3~\mathrm{m~s^{-1}}$, right), at $t=1\,\mathrm{Myr}$. The beam is denoted as an ellipse in the lower left corner of the left panel.}
    \label{fig:model7_high_vfrag}
\end{figure}

Dust traps in the inner disc are expected to retain more dust. This does not hold if dust is fragile and fragments around the trap, as in Model 7. Nevertheless, if dust is less fragile (i.e. has a higher $v_\mathrm{frag}$), larger particles can be formed and trapped effectively around the pressure bump, limiting the supply to the inner disc. The dust concentration associated with the added dust trap therefore becomes visible in the optically thin inner disc. This is illustrated by the synthetic observation of the modified Model 7 (M7$^{\prime}$ in Table \ref{tb:discussion}), with $v_\mathrm{frag}=3~\mathrm{m~s^{-1}}$, which is shown in the right panel of Fig. \ref{fig:model7_high_vfrag}.

\begin{figure}
    \centering
    \includegraphics[width=1.0\linewidth]{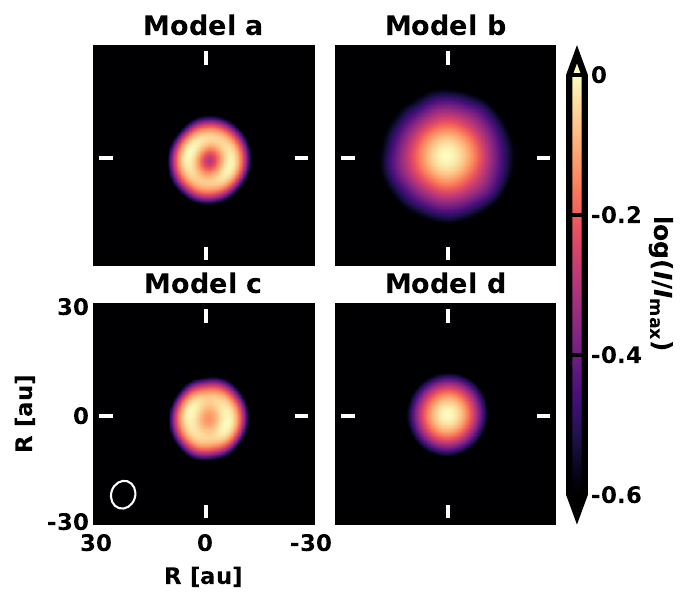}
    \caption{The synthetic ALMA observations at $\lambda = 1.3\,\mathrm{mm}$ for Models a-d in Section \ref{sec:disc_invisble}, at $t=1\,\mathrm{Myr}$. The beam size is plotted as an ellipse in the lower left corner.}
    \label{fig:invis_bump}
\end{figure}

We extend our experiments on the ring at $\sim 6~\mathrm{au}$ to discs with homogeneous $\alpha_\mathrm{SS}$, and find that models with distinct combinations of $v_\mathrm{frag}$ and $\alpha_\mathrm{SS}$ can look similar in synthetic observations\footnote{They may have very different continuum fluxes and measured radii, but these depend strongly on the initial conditions.}. This indicates degeneracy between these two parameters, which arises because a less turbulent disc is more tolerant to the fragility of dust.
This is shown in Fig. \ref{fig:invis_bump}; the parameters and properties of these four models (a--d) are listed in Table \ref{tb:discussion}. All other parameters related to the dust trap are the same as in Model 7, though we note that the planetary mass required to carve an equivalent gap in the gas may differ with the change of $\alpha_\mathrm{SS}$.  

\begin{table*}
\caption{Input parameters and resulting dust disc properties at $\lambda=1.3\,\mathrm{mm}$ for the models discussed in Section \ref{sec:discussion} computed with \textsc{DSHARP} opacities \citep{2018ApJ...869L..45B}.}\label{tb:discussion}
\begin{tabular}{cccccccc}
\hline
\hline
(1)& (2) & (3)& (4) & (5) & (6) & (7) & (8)\\
Section & Model & $\alpha_\mathrm{SS}$  & $v_\mathrm{frag}$ & Porosity ($f$)& $M_\mathrm{d, tot}^\mathrm{1Myr}$ & $F_\mathrm{1.3mm}^\mathrm{1Myr}$ & $R_{d,\mathrm{90\%}}^\mathrm{1Myr}$\\
&&& [$\mathrm{m~s^{-1}}$] & &[$M_\oplus$]  & [mJy] & [au]\\ 
\hline
5.2 & M7$^{\prime}$ & $10^{-4}\rightarrow10^{-3}$ & 3 & 1 & 112.92& 7.7& 25.97\\
5.2 & a & $10^{-4}$ & 6 & 1 &92.01 & 3.34 & 13.73 \\ 
5.2 & b & $10^{-4}$ & 1 & 1 &124.35 & 79.84& 46.81\\
5.2 & c & $4\times10^{-4}$ & 10 &1 & 82.33 & 4.60 & 12.79 \\ 
5.2 & d & $10^{-3}$ & 4 & 1 &24.59 & 27.93 & 25.31 \\ 
5.3.1 & M2$^{\prime}$,1 & $10^{-4}\rightarrow10^{-3}$ & 1 & 0.1 & - & - & -\\
5.3.1 & M2$^{\prime}$,2 & $10^{-4}\rightarrow10^{-3}$ & 1 & 0.05 & - & - & -\\
5.3.2 & M2$^{\prime}$,3 & $10^{-4}\rightarrow10^{-3}$ & 3 & 1 & - & - & -\\
5.3.2 & M2$^{\prime}$,4 & $10^{-4}\rightarrow10^{-3}$ & 5 & 1 & - & - & -\\
5.3.2 & M2$^{\prime}$,5 & $10^{-4}\rightarrow10^{-3}$ & 7 & 1 & - & - & -\\
\hline
\end{tabular}

\footnotesize{\textbf{Note}: Dust total mass, continuum fluxes and observed dust radii are not provided for models requiring additional assumptions on the opacity (M2$^{\prime}$,1 and M2$^{\prime}$,2) and for models lacking effective traps when the fragmentation velocity is high (M2$^{\prime}$,3, M2$^{\prime}$,4 and M2$^{\prime}$,5)}\\
\end{table*}

The poorly resolved ring-like feature associated with the dust trap is visible for Models a and c, but always remains invisible for Models b and d. Models a and c have resilient dust, which grows even in the turbulent inner disc, so large particles can be effectively trapped by the pressure bump. However, the relatively fragile dust in Models b and d (which resembles Model 7) fragments and replenishes the inner disc, leading to an optically thick inner disc.

Models a and c also clearly illustrate that the mm continuum flux can be an unreliable indicator of the actual mass reservoir. Despite their low continuum fluxes and small disc sizes, they have high dust mass, but $>55$ per cent of the mass is stored in particles larger than $1\,\mathrm{cm}$. This offers a potential solution to the mass budget problem, especially for discs with small sizes and low mm fluxes. Multi-wavelength observations to measure the dust size distributions would provide a better estimate of the true mass reservoir in protoplanetary discs \citep[e.g.][]{2023ApJ...942....4X}.

\subsection{Evaluating model robustness}\label{sec:diss_mmbarrier}
The models presented in Section \ref{sec:result} only consider compact dust with low fragmentation velocities. Resilient and fluffy dust, which does not fragment even at collision velocities of tens of $\mathrm{m\,s^{-1}}$ \citep{2015A&A...574A..83K}, can enter the Stokes regime, facilitating dust growth and overcoming the radial drift barrier. In this section, we explore whether the size of a compact disc is still limited by the dead zone if our assumptions about dust compaction and $v_\mathrm{frag}$ are relaxed.

\subsubsection{Porous particles}
The importance of porous dust has been highlighted by recent modelling of multi-wavelength and polarization observations \citep[e.g.][]{2023ApJ...944L..43T, 2023ApJ...953...96Z, 2024A&A...688A..31M, 2024NatAs...8.1148U}. In \textsc{DustPy}, dust porosity is treated simply by lowering the (spherical) monomer density, without accounting for the morphologies of dust aggregates\footnote{This is achieved by multiplying the fill factor $f$ to the monomer (material) density ($\rho_s^{\prime}=f\rho_s$, $f\leq 1$). For a given dust size, a lower monomer density corresponds to a smaller Stokes number ($\mathrm{St}\propto \rho_s^{\prime~2/3}$, see Eq. \ref{eq:stokes}).}. This hinders proper estimation of the porosity, which requires knowledge of the number of monomers in one dust aggregate and the size of the aggregate, in addition to the monomer size. Our adopted (\textsc{DSHARP}) opacities also assume an opacity of zero for the vacuum fraction, leading to unrealistically low continuum fluxes when dust is porous. Additional assumptions are necessary to treat this problem more properly \citep{2024NatAs...8.1148U}. In light of these simplifications, we consider only the theoretical results, without computing synthetic observations, and experiment with only two fill factors, $f=0.1$ and $0.05$, for Model 2. The two models are labelled as M2$^{\prime}$,1 and M2$^{\prime}$, 2 in Table \ref{tb:discussion}, respectively.

\begin{figure}
    \centering
    \includegraphics[width=1.0\linewidth]{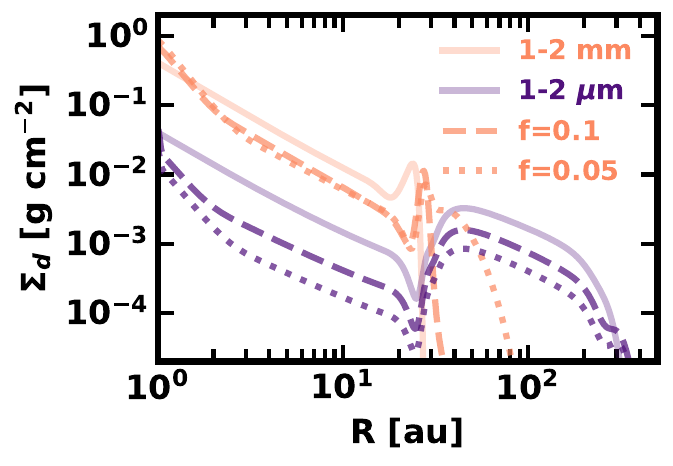}
    \caption{The integrated surface density of $1$-$2~\mathrm{\mu m}$ (purple) and $1$-$2~\mathrm{mm}$ (orange) dust for Model 2 with compact dust (f=1, solid lines), and for modified Model 2 with porous dust (M2$^{\prime}$,1: f=0.1, dashed lines; M2$^{\prime},2$: f=0.05, dotted lines).}
    \label{fig:porous_dust}
\end{figure}

Fig. \ref{fig:porous_dust} shows that the integrated surface density of both micron and mm grains at a few au is lower for porous dust than for compact dust, as increased porosity results in larger particles across the entire disc. However, the radial extent of the mm-dust disc does not change unless dust is very porous ($f=0.05$). Such high porosity is less likely for mm-dust \citep[e.g.][]{2019ApJ...885...52T}, implying that increased porosity alone is unlikely to result in a mm disc that is larger than the MRI-suppressed region. However, a more realistic treatment of dust porosity and the dust opacity is needed to understand this problem fully.

\subsubsection{Fragmentation velocity}

Resilient dust shifts discs into a drift-dominated regime, where larger dust grains are less prone to fragmentation but are lost more rapidly without effective traps. In the absence of such traps, we apply a series of higher fragmentation velocities (3/5/7 $\mathrm{m~s^{-1}}$) to Model 2 to examine the formation of mm-dust in MRI-active regions. We measure the integrated surface density for grain sizes of $0.1$-$1~\mathrm{cm}$ in numerical models at $0.5~\mathrm{Myr}$, before the majority of the dust drifts inwards.

\begin{figure}
    \centering
    \includegraphics[width=1.0\linewidth]{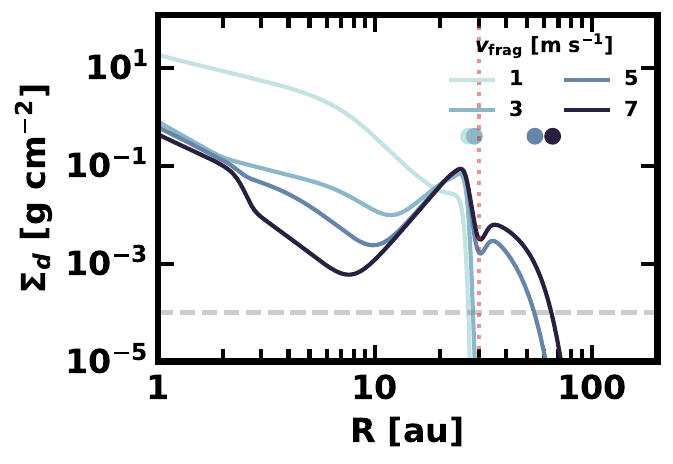}
    \caption{The integrated surface density of $0.1$-$1~\mathrm{cm}$ dust for (modified) Model 2 when the fragmentation velocity is set to be 1, 3, 5, 7 $\mathrm{m~s^{-1}}$ at $0.5~\mathrm{Myr}$ (darker colours for higher velocities). Filled circles with corresponding colours show the radial extent of disc sizes cut by a threshold of $10^{-4}~\mathrm{g~cm^{-2}}$. The pink dotted line indicates the transition from MRI-inactive to MRI-active regions.} 
    \label{fig:vfrag_R}
\end{figure}

Fig. \ref{fig:vfrag_R} shows that mm dust can form beyond the dead zone only when $v_\mathrm{frag}\gtrsim 5~\mathrm{m~s^{-1}}$. The more extended disc beyond $R_t=30~\mathrm{au}$ in Fig. \ref{fig:vfrag_R} emits only weakly, comparable to the noise level, and does not change the observed disc size significantly. In $1.3$-mm synthetic observations: Model M2$^{\prime}$,5 ($v_\mathrm{frag}=7~\mathrm{m~s^{-1}}$) has  $R_{d,\mathrm{90\%}}=29.46~\mathrm{au}$ at $0.5~\mathrm{Myr}$, compared to $R_{d,\mathrm{90\%}}=26.67~\mathrm{au}$ for Model 2 (at $1~\mathrm{Myr}$). This implies that a higher fragmentation velocity is also not sufficient to allow mm-dust discs to extend significantly beyond the outer edge of the dead zone. A proper consideration of pressure bumps in discs in future work could provide deeper understanding of this problem.

\subsubsection{Decoupling of the different turbulence parameters}\label{sec:disc_var_alpha}

Model 3 shows that it is difficult to form mm-size dust in the MRI-active region even with turbulence as weak as $\alpha_{\mathrm {SS}} = 5\times10^{-4}$, below what is typically inferred from observations of outer discs \citep{2023NewAR..9601674R}. Nevertheless, current measurements of the turbulence parameter $\alpha$ may measure velocity perturbations arising from various sources. We assume that our models share the same values for $\alpha_\mathrm{SS}$, $\delta_\mathrm{vert}$, $\delta_\mathrm{turb}$ and $\delta_r$, but this may not always be the case. 

Turbulence $\alpha$ measured from line broadening traces the gas transport $\alpha_\mathrm{SS}$, but may also be sensitive to other physical mechanisms, such as vertical shear instabilities, gravitational instabilities, and bulk motions induced by the stellar environment. These may be particularly relevant for radially extended discs which show higher than usual turbulence, such as IM Lup \citep{2024MNRAS.532..363F} and DM Tau \citep{2020ApJ...895..109F}. 

By contrast, turbulence derived from the dust scale-height \citep[e.g.][]{2021ApJ...912..164D}, or the width of dust rings \citep[e.g.][]{2018ApJ...869L..46D} probes the \textit{local} values of $\delta_\mathrm{vert}$ (see Eq. \ref{eq:dust_sh}) and $\delta_r$ (see Eq. \ref{eq:diff}), respectively. Future work measuring turbulence for individual targets through multiple methods is needed to provide a more comprehensive understanding. If the fragmentation velocity is truly as low as laboratory experiments suggest \citep[e.g.][]{2019ApJ...873...58M}, the low turbulence required for mm dust formation is difficult to reconcile with the higher levels of turbulence required to explain observed stellar accretion rates. This may suggest that additional mechanisms, such as magnetized winds \citep[e.g.][]{1982MNRAS.199..883B, 2013ApJ...769...76B}, are needed to drive protoplanetary disc accretion.

\subsection{Diverse compact discs}

Our results suggest that dead zones can explain some of the observed compact discs, but compact discs may also form through other physical mechanisms. By considering multiple observables, such as gas-to-dust size ratios, the shape of radial intensity profiles, and dust disc sizes across multi-wavelength observations, we can potentially distinguish between different formation mechanisms for compact discs (or at least rule out some possibilities).

Discs which are born small are predicted to have gas-to-dust size ratios comparable to radially extended and substructured discs around single stars. By contrast, small discs caused by tidal truncation are found to have a larger size ratio ($R_{g,\mathrm{95\%}}/R_{d,\mathrm{95\%}}$) \citep{2022A&A...662A.121R} than isolated systems. This may be related to enhanced radial drift in binary systems, giving rise to a smaller dust disc when no dust traps are present \citep[e.g.][]{2021MNRAS.504.2235Z}. Large size ratios are also expected for drift-dominated \citep{2019A&A...626L...2F} and fragmentation-limited discs (Section \ref{sec:disc_SLR}). Multi-wavelength observations may help to differentiate the latter two mechanisms. Drift-dominated discs are predicted to have smaller sizes at longer observing wavelengths, while fragmentation-dominated discs are predicted to have slightly smaller (or even similar) discs sizes at $3$-mm than at $1.3$-mm when the transition from the MRI-inactive to the MRI-active region is sharp (see Table \ref{tb:models1} and \ref{tb:models3}). We note that the effects of optical depth can also contribute to the gas-to-dust size ratio \citep[e.g.][]{2017A&A...605A..16F}, but this is not considered here.

The sharpness of the outer edge of the radial intensity profile can also be used as a discriminator for compact discs. In addition to sharper edges observed for discs in multiple systems than in isolated systems, caused by tidal truncation \citep{2019A&A...628A..95M}, varying sharpness of the outer edge has been identified in single star systems \citep[e.g.][]{2021A&A...649A.122P,2023ApJ...952..108Z,2024ApJ...966...59S}. \citet{2023A&A...673A..77R} and \citet{2024A&A...682A..55M} also observed weak ring-like emission at the outer edge, which is strikingly similar to the feature caused by `hitchhiking' effects introduced in Section \ref{sec:result_DZ_revis} (see also Fig. \ref{fig:radial_M7}). Analysis of radial intensity profiles also demonstrates that compact discs are not simply scaled-down versions of radially extended substructured discs \citep[e.g. CX Tau, Sz 65, Sz 66 and compact discs in DSHARP samples,][]{2019A&A...626L...2F, 2024A&A...682A..55M}. It remains unclear whether substructured discs can be formed from initially small discs which trap dust from late infall at larger radii \citep[e.g.][]{2023EPJP..138..272K, kuznetsova2022anisotropic}. 

In addition to physical quantities, chemical information obtained from both ALMA and JWST may provide hints on the formation mechanisms of compact discs. \citet{2021A&A...651A..48M} invoked compact discs to explain discs that are detected in continuum but that are faint or not detected in CO isotopologues, aligning with ideas proposed by \citet{2016ApJ...827..142B} and observations by \citet{2014A&A...564A..95P}. Drift-dominated discs are predicted to have higher O/H ratio in the inner disc due to the sublimation of water-rich ices carried by the rapid drifting pebbles after crossing the snow line \citep{2006Icar..181..178C}. This has been corroborated by high-resolution spectroscopy from JWST \citep[e.g.][]{2023ApJ...957L..22B}, but the detectability of $\mathrm{H_2O}$-rich inner disc might be complicated by disc evolution and other processes \citep[e.g.][]{2024arXiv241201895S, vlasblom2024}. All of these studies have been limited to a few targets. Systematic studies based on larger samples will draw firmer conclusions and provide greater insight.

\subsection{Limitations}
The evolution of viscous discs in this study does not include the magnetised wind, which is thought to dominate angular momentum transport, at least in some parts of the disc. Magnetised winds in 1-D models introduce an inward advection term and a mass-loss term \citep[e.g.][]{2016A&A...596A..74S,2022MNRAS.512.2290T}. The advection term hastens inward drift of dust when no effective dust traps are present. The enhancement in the radial velocity depends on the efficiency of angular momentum transported by the wind. Whether the gas disc size decreases significantly, which will further affect the dust disc size, depends on the efficiency of viscosity relative to that of the wind. This question, while relevant, is beyond the scope of this study. We note, however, when the wind is strong, dust can be susceptible to destructive collisions. Dust can also be removed by the wind when dust entrainment is accounted, but the reduced dust mass by this process should be negligible. Simulations show that only (sub)micron-sized dust concentrated in the inner disc above the wind base \citep[e.g.][]{2022A&A...659A..42R}, which accounts for a very small fraction of the total dust mass, can be entrained in the wind. The increased dust-to-gas ratio may benefit planet formation, but to what degree this ratio is enhanced also depends on the lever arm $\lambda$, which is estimated to be $\sim 2$ \citep[e.g.][]{2018A&A...618A.120L, 2020A&A...634L..12D} or higher ($\sim 4$-$5$) from observations \citep[e.g.][]{2017A&A...607L...6T, 2021ApJ...907L..41L, 2025arXiv250103920B}. The effects of the lever arm on long-term gas disc evolution have been studied in \citet{2024_disc_evo}.

Our models simply treat the dead zone as a constant two-zone model, and naively apply various dead zone models to discs with identical physical properties. In more realistic calculations, the shape of dead zones and the $\alpha_\mathrm{SS}$ parameter in the MRI-suppressed and MRI-active regions will represent discs with various initial conditions, and will co-evolve with dust grains, ionization, and magnetic fields. We may therefore see different dead zone profiles in discs with different physical properties, which cannot be captured by the simplified models presented here. 

In addition, the fragmentation velocity is treated as a global constant throughout this paper, but it can be radius-dependent. This does not necessarily relate to the variation in the fragmentation velocity around the icelines \citep[e.g.][]{2017ApJ...845...68P, 2020MNRAS.493.1013A}, but can arise from other physico-chemical effects. Furthermore, there may even be a range of fragmentation velocities at a given radius. 

Finally, we note that our models do not account for the back-reaction of the dust on the gas \citep[e.g.][]{2017ApJ...844..142K, 2018MNRAS.479.4187D, 2020A&A...635A.149G}. This should be negligible in models without dust traps, but where traps exist, the high local dust-to-gas ratio can perturb gas motion significantly, with knock-on effects on the dust dynamics. 

\section{Conclusion}\label{sec:conclusion}
In this paper we have proposed that the interaction between fragile dust and moderate turbulence, as expected in MRI-active regions, provides an alternative mechanism to produce radially compact protoplanetary discs which are consistent with current observations. We ran a suite of 1-D gas and dust evolution models in \textsc{DustPy} to test this scenario, and generated synthetic observations from numerical models to compare with real observations. We consider a variety of models with radially varying $\alpha_\mathrm{SS}(R)$, and explore a wide range of dust properties. Our main results are as follows:
\begin{itemize}
   \item A radially compact protoplanetary disc can be naturally created when fragile dust ($v_{\mathrm {frag}} \simeq 1$\,m\,s$^{-1}$) is destroyed in a moderately turbulent environment ($\alpha_\mathrm{SS}=10^{-3}$), which can be the case in MRI-active regions beyond the dead zone.
   \item In this scenario, the observed mm-dust disc size of compact discs is determined by the radial extent of the dead zone. The observed disc sizes at $1.3$ mm and $3.0$ mm are similar, which is consistent with ALMA multi-wavelength observations, when the transition of the gas transport parameter $\alpha_\mathrm{SS}$ is sharp (i.e. the dead zone has a sharp outer edge). 
    \item Pressure bumps present in the inner disc ($\sim 6~\mathrm{au}$) are not effective dust traps. Dust fragments around the bumps and replenishes the inner disc. This yields an optically thick inner disc, which hides the ring-like feature in dust emission from observations.
    \item A sharp drop in the gas surface density can hasten the radial drift of $\lesssim$ mm dust locally. We refer to this process as the `hitchhiking' effect. This is a case we expect in dead zones with sharp transitions from the MRI-inactive to MRI-active regions. The accumulated dust at the outer edge of these dead zones can create a low-contrast ring-like feature in observations with high resolution ($0.\arcsec02$). This feature is very similar to those observed in compact discs such as Sz 66 \citep{2024A&A...682A..55M} and MP Mus \citep{2023A&A...673A..77R}.
    \item Compact discs formed in this scenario follow the size-luminosity relation that is commonly found in other studies ($R_\mathrm{eff}\propto L_\mathrm{mm}^{0.5}$). Our models also predict a shallower relation at later evolutionary stages, as seen in older star-forming regions such as Upper Sco. 
    \item Increasing the dust porosity, or considering less fragile dust (i.e. increasing the fragmentation velocity), does not significantly increase the observed size of compact discs formed by this mechanism.
    \item We have identified significant degeneracy between the gas viscous transport parameter $\alpha_\mathrm{SS}$ and the fragmentation velocity $v_\mathrm{frag}$. Generally, a less turbulent environment is more tolerant to fragile dust.
\end{itemize}

\section*{Acknowledgements}
ST acknowledges the University of Leicester for a Future 100 Studentship. RA acknowledges funding from the Science \& Technology Facilities Council (STFC) through Consolidated Grant ST/W000857/1. ST would like to thank the organizers, lecturers, and participants of the Dustbusters School II for sparking her interest in compact protoplanetary discs. ST also thanks Phil Armitage for his warm hospitality during her visit to the State University of New York, Stony Brook. This project has received funding from the European Union’s Horizon 2020 research and innovation programme under the Marie Skłodowska-Curie grant agreement No 823823 (DUSTBUSTERS). This research used the ALICE High Performance Computing Facility at the University of Leicester. 
For the purpose of open access, the author has applied a Creative Commons Attribution (CC BY) licence to the Author Accepted Manuscript version arising from this submission.

\section*{Data Availability}

Data generated in simulations and codes reproducing figures in this work are available on reasonable request to the corresponding author. 
This work made use of \textsc{DustPy} \citep{2022ApJ...935...35S}, \textsc{DustyPyLib} \citep{2023ascl.soft10005S}, \textsc{RADMC-3D} \citep{2012ascl.soft02015D}, \textsc{Jupyter} \citep{jupyter}, \textsc{Matplotlib} \citep{matplotlib}, \textsc{Numpy} \citep{numpy, numpy2}, \textsc{Scipy} \citep{scipy2020}, \textsc{Astropy} \citep{astropy2018} and \textsc{Pandas} \citep{pandas2010, pandas2020}.



\bibliographystyle{mnras}
\bibliography{reference}

\begin{thebibliography}{}
\makeatletter
\relax
\def\mn@urlcharsother{\let\do\@makeother \do\$\do\&\do\#\do\^\do\_\do\%\do\~}
\def\mn@doi{\begingroup\mn@urlcharsother \@ifnextchar [ {\mn@doi@} {\mn@doi@[]}}
\def\mn@doi@[#1]#2{\def\@tempa{#1}\ifx\@tempa\@empty \href {http://dx.doi.org/#2} {doi:#2}\else \href {http://dx.doi.org/#2} {#1}\fi \endgroup}
\def\mn@eprint#1#2{\mn@eprint@#1:#2::\@nil}
\def\mn@eprint@arXiv#1{\href {http://arxiv.org/abs/#1} {{\tt arXiv:#1}}}
\def\mn@eprint@dblp#1{\href {http://dblp.uni-trier.de/rec/bibtex/#1.xml} {dblp:#1}}
\def\mn@eprint@#1:#2:#3:#4\@nil{\def\@tempa {#1}\def\@tempb {#2}\def\@tempc {#3}\ifx \@tempc \@empty \let \@tempc \@tempb \let \@tempb \@tempa \fi \ifx \@tempb \@empty \def\@tempb {arXiv}\fi \@ifundefined {mn@eprint@\@tempb}{\@tempb:\@tempc}{\expandafter \expandafter \csname mn@eprint@\@tempb\endcsname \expandafter{\@tempc}}}

\bibitem[\protect\citeauthoryear{{Alessi}, {Pudritz}  \& {Cridland}}{{Alessi} et~al.}{2020}]{2020MNRAS.493.1013A}
{Alessi} M.,  {Pudritz} R.~E.,   {Cridland} A.~J.,  2020, \mn@doi [\mnras] {10.1093/mnras/staa308}, \href {https://ui.adsabs.harvard.edu/abs/2020MNRAS.493.1013A} {493, 1013}

\bibitem[\protect\citeauthoryear{{Anderson}, {Williams}, {van der Marel}, {Law}, {Ricci}, {Tobin}  \& {Tong}}{{Anderson} et~al.}{2022}]{2022ApJ93855A}
{Anderson} A.~R.,  {Williams} J.~P.,  {van der Marel} N.,  {Law} C.~J.,  {Ricci} L.,  {Tobin} J.~J.,   {Tong} S.,  2022, \mn@doi [\apj] {10.3847/1538-4357/ac8ff0}, \href {https://ui.adsabs.harvard.edu/abs/2022ApJ...938...55A} {938, 55}

\bibitem[\protect\citeauthoryear{{Andrews}, {Terrell}, {Tripathi}, {Ansdell}, {Williams}  \& {Wilner}}{{Andrews} et~al.}{2018}]{2018ApJ...865..157A}
{Andrews} S.~M.,  {Terrell} M.,  {Tripathi} A.,  {Ansdell} M.,  {Williams} J.~P.,   {Wilner} D.~J.,  2018, \mn@doi [\apj] {10.3847/1538-4357/aadd9f}, \href {https://ui.adsabs.harvard.edu/abs/2018ApJ...865..157A} {865, 157}

\bibitem[\protect\citeauthoryear{{Ansdell} et~al.,}{{Ansdell} et~al.}{2016}]{2016ApJ...828...46A}
{Ansdell} M.,  et~al., 2016, \mn@doi [\apj] {10.3847/0004-637X/828/1/46}, \href {https://ui.adsabs.harvard.edu/abs/2016ApJ...828...46A} {828, 46}

\bibitem[\protect\citeauthoryear{{Ansdell}, {Williams}, {Manara}, {Miotello}, {Facchini}, {van der Marel}, {Testi}  \& {van Dishoeck}}{{Ansdell} et~al.}{2017}]{2017AJ....153..240A}
{Ansdell} M.,  {Williams} J.~P.,  {Manara} C.~F.,  {Miotello} A.,  {Facchini} S.,  {van der Marel} N.,  {Testi} L.,   {van Dishoeck} E.~F.,  2017, \mn@doi [\aj] {10.3847/1538-3881/aa69c0}, \href {https://ui.adsabs.harvard.edu/abs/2017AJ....153..240A} {153, 240}

\bibitem[\protect\citeauthoryear{{Arlt} \& {Urpin}}{{Arlt} \& {Urpin}}{2004}]{2004A&A...426..755A}
{Arlt} R.,  {Urpin} V.,  2004, \mn@doi [\aap] {10.1051/0004-6361:20035896}, \href {https://ui.adsabs.harvard.edu/abs/2004A&A...426..755A} {426, 755}

\bibitem[\protect\citeauthoryear{{Bacciotti} et~al.,}{{Bacciotti} et~al.}{2025}]{2025arXiv250103920B}
{Bacciotti} F.,  et~al., 2025, arXiv e-prints, \href {https://ui.adsabs.harvard.edu/abs/2025arXiv250103920B} {p. arXiv:2501.03920}

\bibitem[\protect\citeauthoryear{{Bai}}{{Bai}}{2015}]{2015ApJ...798...84B}
{Bai} X.-N.,  2015, \mn@doi [\apj] {10.1088/0004-637X/798/2/84}, \href {https://ui.adsabs.harvard.edu/abs/2015ApJ...798...84B} {798, 84}

\bibitem[\protect\citeauthoryear{{Bai} \& {Stone}}{{Bai} \& {Stone}}{2013}]{2013ApJ...769...76B}
{Bai} X.-N.,  {Stone} J.~M.,  2013, \mn@doi [\apj] {10.1088/0004-637X/769/1/76}, \href {https://ui.adsabs.harvard.edu/abs/2013ApJ...769...76B} {769, 76}

\bibitem[\protect\citeauthoryear{{Bai}, {Ye}, {Goodman}  \& {Yuan}}{{Bai} et~al.}{2016}]{2016ApJ...818..152B}
{Bai} X.-N.,  {Ye} J.,  {Goodman} J.,   {Yuan} F.,  2016, \mn@doi [\apj] {10.3847/0004-637X/818/2/152}, \href {https://ui.adsabs.harvard.edu/abs/2016ApJ...818..152B} {818, 152}

\bibitem[\protect\citeauthoryear{Balbus \& Hawley}{Balbus \& Hawley}{1991}]{1991ApJ...376..214B}
Balbus S.~A.,  Hawley J.~F.,  1991, \mn@doi [\apj] {10.1086/170270}, 376, 214

\bibitem[\protect\citeauthoryear{{Banzatti} et~al.,}{{Banzatti} et~al.}{2023}]{2023ApJ...957L..22B}
{Banzatti} A.,  et~al., 2023, \mn@doi [\apjl] {10.3847/2041-8213/acf5ec}, \href {https://ui.adsabs.harvard.edu/abs/2023ApJ...957L..22B} {957, L22}

\bibitem[\protect\citeauthoryear{{Barenfeld}, {Carpenter}, {Ricci}  \& {Isella}}{{Barenfeld} et~al.}{2016}]{2016ApJ...827..142B}
{Barenfeld} S.~A.,  {Carpenter} J.~M.,  {Ricci} L.,   {Isella} A.,  2016, \mn@doi [\apj] {10.3847/0004-637X/827/2/142}, \href {https://ui.adsabs.harvard.edu/abs/2016ApJ...827..142B} {827, 142}

\bibitem[\protect\citeauthoryear{{B{\'e}thune}, {Lesur}  \& {Ferreira}}{{B{\'e}thune} et~al.}{2017}]{2017A&A...600A..75B}
{B{\'e}thune} W.,  {Lesur} G.,   {Ferreira} J.,  2017, \mn@doi [\aap] {10.1051/0004-6361/201630056}, \href {https://ui.adsabs.harvard.edu/abs/2017A&A...600A..75B} {600, A75}

\bibitem[\protect\citeauthoryear{{Birnstiel}, {Dullemond}  \& {Brauer}}{{Birnstiel} et~al.}{2009}]{2009A&A...503L...5B}
{Birnstiel} T.,  {Dullemond} C.~P.,   {Brauer} F.,  2009, \mn@doi [\aap] {10.1051/0004-6361/200912452}, \href {https://ui.adsabs.harvard.edu/abs/2009A&A...503L...5B} {503, L5}

\bibitem[\protect\citeauthoryear{Birnstiel, Dullemond  \& Brauer}{Birnstiel et~al.}{2010}]{2010A&A...513A..79B}
Birnstiel T.,  Dullemond C.,   Brauer F.,  2010, \mn@doi [\aap] {10.1051/0004-6361/200913731}, 513, A79

\bibitem[\protect\citeauthoryear{{Birnstiel}, {Klahr}  \& {Ercolano}}{{Birnstiel} et~al.}{2012}]{2012A&A...539A.148B}
{Birnstiel} T.,  {Klahr} H.,   {Ercolano} B.,  2012, \mn@doi [\aap] {10.1051/0004-6361/201118136}, \href {https://ui.adsabs.harvard.edu/abs/2012A&A...539A.148B} {539, A148}

\bibitem[\protect\citeauthoryear{{Birnstiel} et~al.,}{{Birnstiel} et~al.}{2018}]{2018ApJ...869L..45B}
{Birnstiel} T.,  et~al., 2018, \mn@doi [\apjl] {10.3847/2041-8213/aaf743}, \href {https://ui.adsabs.harvard.edu/abs/2018ApJ...869L..45B} {869, L45}

\bibitem[\protect\citeauthoryear{{Blandford} \& {Payne}}{{Blandford} \& {Payne}}{1982}]{1982MNRAS.199..883B}
{Blandford} R.~D.,  {Payne} D.~G.,  1982, \mn@doi [\mnras] {10.1093/mnras/199.4.883}, \href {https://ui.adsabs.harvard.edu/abs/1982MNRAS.199..883B} {199, 883}

\bibitem[\protect\citeauthoryear{{Blum} \& {Wurm}}{{Blum} \& {Wurm}}{2008}]{2008ARA&A..46...21B}
{Blum} J.,  {Wurm} G.,  2008, \mn@doi [\araa] {10.1146/annurev.astro.46.060407.145152}, \href {https://ui.adsabs.harvard.edu/abs/2008ARA&A..46...21B} {46, 21}

\bibitem[\protect\citeauthoryear{{CASA Team} et~al.,}{{CASA Team} et~al.}{2022}]{2022PASP..134k4501C}
{CASA Team} et~al., 2022, \mn@doi [\pasp] {10.1088/1538-3873/ac9642}, \href {https://ui.adsabs.harvard.edu/abs/2022PASP..134k4501C} {134, 114501}

\bibitem[\protect\citeauthoryear{{Cazzoletti} et~al.,}{{Cazzoletti} et~al.}{2019}]{2019A&A...626A..11C}
{Cazzoletti} P.,  et~al., 2019, \mn@doi [\aap] {10.1051/0004-6361/201935273}, \href {https://ui.adsabs.harvard.edu/abs/2019A&A...626A..11C} {626, A11}

\bibitem[\protect\citeauthoryear{{Ciesla} \& {Cuzzi}}{{Ciesla} \& {Cuzzi}}{2006}]{2006Icar..181..178C}
{Ciesla} F.~J.,  {Cuzzi} J.~N.,  2006, \mn@doi [\icarus] {10.1016/j.icarus.2005.11.009}, \href {https://ui.adsabs.harvard.edu/abs/2006Icar..181..178C} {181, 178}

\bibitem[\protect\citeauthoryear{Cieza et~al.,}{Cieza et~al.}{2019}]{2019MNRAS.482..698C}
Cieza L.~A.,  et~al., 2019, \mn@doi [\mnras] {10.1093/mnras/sty2653}, 482, 698

\bibitem[\protect\citeauthoryear{{Cox} et~al.,}{{Cox} et~al.}{2017}]{2017ApJ...851...83C}
{Cox} E.~G.,  et~al., 2017, \mn@doi [\apj] {10.3847/1538-4357/aa97e2}, \href {https://ui.adsabs.harvard.edu/abs/2017ApJ...851...83C} {851, 83}

\bibitem[\protect\citeauthoryear{{Delage}, {Okuzumi}, {Flock}, {Pinilla}  \& {Dzyurkevich}}{{Delage} et~al.}{2022}]{2022A&A...658A..97D}
{Delage} T.~N.,  {Okuzumi} S.,  {Flock} M.,  {Pinilla} P.,   {Dzyurkevich} N.,  2022, \mn@doi [\aap] {10.1051/0004-6361/202141689}, \href {https://ui.adsabs.harvard.edu/abs/2022A&A...658A..97D} {658, A97}

\bibitem[\protect\citeauthoryear{{Delage}, {G{\'a}rate}, {Okuzumi}, {Yang}, {Pinilla}, {Flock}, {Stammler}  \& {Birnstiel}}{{Delage} et~al.}{2023}]{2023A&A...674A.190D}
{Delage} T.~N.,  {G{\'a}rate} M.,  {Okuzumi} S.,  {Yang} C.-C.,  {Pinilla} P.,  {Flock} M.,  {Stammler} S.~M.,   {Birnstiel} T.,  2023, \mn@doi [\aap] {10.1051/0004-6361/202244731}, \href {https://ui.adsabs.harvard.edu/abs/2023A&A...674A.190D} {674, A190}

\bibitem[\protect\citeauthoryear{{Delussu}, {Birnstiel}, {Miotello}, {Pinilla}, {Rosotti}  \& {Andrews}}{{Delussu} et~al.}{2024}]{2024A&A...688A..81D}
{Delussu} L.,  {Birnstiel} T.,  {Miotello} A.,  {Pinilla} P.,  {Rosotti} G.,   {Andrews} S.~M.,  2024, \mn@doi [\aap] {10.1051/0004-6361/202450328}, \href {https://ui.adsabs.harvard.edu/abs/2024A&A...688A..81D} {688, A81}

\bibitem[\protect\citeauthoryear{Dipierro, Laibe, Alexander  \& Hutchison}{Dipierro et~al.}{2018}]{2018MNRAS.479.4187D}
Dipierro G.,  Laibe G.,  Alexander R.,   Hutchison M.,  2018, \mn@doi [\mnras] {10.1093/mnras/sty1701}, 479, 4187

\bibitem[\protect\citeauthoryear{{Doi} \& {Kataoka}}{{Doi} \& {Kataoka}}{2021}]{2021ApJ...912..164D}
{Doi} K.,  {Kataoka} A.,  2021, \mn@doi [\apj] {10.3847/1538-4357/abe5a6}, \href {https://ui.adsabs.harvard.edu/abs/2021ApJ...912..164D} {912, 164}

\bibitem[\protect\citeauthoryear{{Dong} et~al.,}{{Dong} et~al.}{2018}]{2018ApJ...860..124D}
{Dong} R.,  et~al., 2018, \mn@doi [\apj] {10.3847/1538-4357/aac6cb}, \href {https://ui.adsabs.harvard.edu/abs/2018ApJ...860..124D} {860, 124}

\bibitem[\protect\citeauthoryear{{Draine}}{{Draine}}{2003}]{2003ARA&A..41..241D}
{Draine} B.~T.,  2003, \mn@doi [\araa] {10.1146/annurev.astro.41.011802.094840}, \href {https://ui.adsabs.harvard.edu/abs/2003ARA&A..41..241D} {41, 241}

\bibitem[\protect\citeauthoryear{{Dr{\k{a}}{\.z}kowska}, {Windmark}  \& {Dullemond}}{{Dr{\k{a}}{\.z}kowska} et~al.}{2014}]{2014A&A...567A..38D}
{Dr{\k{a}}{\.z}kowska} J.,  {Windmark} F.,   {Dullemond} C.~P.,  2014, \mn@doi [\aap] {10.1051/0004-6361/201423708}, \href {https://ui.adsabs.harvard.edu/abs/2014A&A...567A..38D} {567, A38}

\bibitem[\protect\citeauthoryear{{Dubrulle}, {Morfill}  \& {Sterzik}}{{Dubrulle} et~al.}{1995}]{1995Icar..114..237D}
{Dubrulle} B.,  {Morfill} G.,   {Sterzik} M.,  1995, \mn@doi [\icarus] {10.1006/icar.1995.1058}, \href {https://ui.adsabs.harvard.edu/abs/1995Icar..114..237D} {114, 237}

\bibitem[\protect\citeauthoryear{{Dullemond}, {Juhasz}, {Pohl}, {Sereshti}, {Shetty}, {Peters}, {Commercon}  \& {Flock}}{{Dullemond} et~al.}{2012}]{2012ascl.soft02015D}
{Dullemond} C.~P.,  {Juhasz} A.,  {Pohl} A.,  {Sereshti} F.,  {Shetty} R.,  {Peters} T.,  {Commercon} B.,   {Flock} M.,  2012, {RADMC-3D: A multi-purpose radiative transfer tool}, Astrophysics Source Code Library, record ascl:1202.015

\bibitem[\protect\citeauthoryear{{Dullemond} et~al.,}{{Dullemond} et~al.}{2018}]{2018ApJ...869L..46D}
{Dullemond} C.~P.,  et~al., 2018, \mn@doi [\apjl] {10.3847/2041-8213/aaf742}, \href {https://ui.adsabs.harvard.edu/abs/2018ApJ...869L..46D} {869, L46}

\bibitem[\protect\citeauthoryear{{Dzyurkevich}, {Turner}, {Henning}  \& {Kley}}{{Dzyurkevich} et~al.}{2013}]{2013ApJ...765..114D}
{Dzyurkevich} N.,  {Turner} N.~J.,  {Henning} T.,   {Kley} W.,  2013, \mn@doi [\apj] {10.1088/0004-637X/765/2/114}, \href {https://ui.adsabs.harvard.edu/abs/2013ApJ...765..114D} {765, 114}

\bibitem[\protect\citeauthoryear{Eisner et~al.,}{Eisner et~al.}{2018}]{2018ApJ...860...77E}
Eisner J.,  et~al., 2018, \mn@doi [\apj] {10.3847/1538-4357/aac3e2}, 860, 77

\bibitem[\protect\citeauthoryear{{Facchini}, {Birnstiel}, {Bruderer}  \& {van Dishoeck}}{{Facchini} et~al.}{2017}]{2017A&A...605A..16F}
{Facchini} S.,  {Birnstiel} T.,  {Bruderer} S.,   {van Dishoeck} E.~F.,  2017, \mn@doi [\aap] {10.1051/0004-6361/201630329}, \href {https://ui.adsabs.harvard.edu/abs/2017A&A...605A..16F} {605, A16}

\bibitem[\protect\citeauthoryear{Facchini et~al.,}{Facchini et~al.}{2019}]{2019A&A...626L...2F}
Facchini S.,  et~al., 2019, \mn@doi [\aap] {10.1051/0004-6361/201935496}, 626, L2

\bibitem[\protect\citeauthoryear{Flaherty et~al.,}{Flaherty et~al.}{2020}]{2020ApJ...895..109F}
Flaherty K.,  et~al., 2020, \mn@doi [\apj] {10.3847/1538-4357/ab8cc5}, 895, 109

\bibitem[\protect\citeauthoryear{{Flaherty} et~al.,}{{Flaherty} et~al.}{2024}]{2024MNRAS.532..363F}
{Flaherty} K.,  et~al., 2024, \mn@doi [\mnras] {10.1093/mnras/stae1480}, \href {https://ui.adsabs.harvard.edu/abs/2024MNRAS.532..363F} {532, 363}

\bibitem[\protect\citeauthoryear{Flock, Fromang, Turner  \& Benisty}{Flock et~al.}{2016}]{2016ApJ...827..144F}
Flock M.,  Fromang S.,  Turner N.,   Benisty M.,  2016, \mn@doi [\apj] {10.3847/0004-637X/827/2/144}, 827, 144

\bibitem[\protect\citeauthoryear{{Flock}, {Turner}, {Mulders}, {Hasegawa}, {Nelson}  \& {Bitsch}}{{Flock} et~al.}{2019}]{2019A&A...630A.147F}
{Flock} M.,  {Turner} N.~J.,  {Mulders} G.~D.,  {Hasegawa} Y.,  {Nelson} R.~P.,   {Bitsch} B.,  2019, \mn@doi [\aap] {10.1051/0004-6361/201935806}, \href {https://ui.adsabs.harvard.edu/abs/2019A&A...630A.147F} {630, A147}

\bibitem[\protect\citeauthoryear{{Gammie}}{{Gammie}}{1996}]{1996ApJ...457..355G}
{Gammie} C.~F.,  1996, \mn@doi [\apj] {10.1086/176735}, \href {https://ui.adsabs.harvard.edu/abs/1996ApJ...457..355G} {457, 355}

\bibitem[\protect\citeauthoryear{{G{\'a}rate}, {Birnstiel}, {Dr{\k{a}}{\.z}kowska}  \& {Stammler}}{{G{\'a}rate} et~al.}{2020}]{2020A&A...635A.149G}
{G{\'a}rate} M.,  {Birnstiel} T.,  {Dr{\k{a}}{\.z}kowska} J.,   {Stammler} S.~M.,  2020, \mn@doi [\aap] {10.1051/0004-6361/201936067}, \href {https://ui.adsabs.harvard.edu/abs/2020A&A...635A.149G} {635, A149}

\bibitem[\protect\citeauthoryear{{G{\'a}rate} et~al.,}{{G{\'a}rate} et~al.}{2021}]{2021A&A...655A..18G}
{G{\'a}rate} M.,  et~al., 2021, \mn@doi [\aap] {10.1051/0004-6361/202141444}, \href {https://ui.adsabs.harvard.edu/abs/2021A&A...655A..18G} {655, A18}

\bibitem[\protect\citeauthoryear{{Gundlach} \& {Blum}}{{Gundlach} \& {Blum}}{2015}]{2015ApJ...798...34G}
{Gundlach} B.,  {Blum} J.,  2015, \mn@doi [\apj] {10.1088/0004-637X/798/1/34}, \href {https://ui.adsabs.harvard.edu/abs/2015ApJ...798...34G} {798, 34}

\bibitem[\protect\citeauthoryear{{Harris} et~al.,}{{Harris} et~al.}{2020}]{numpy2}
{Harris} C.~R.,  et~al., 2020, \mn@doi [\nat] {10.1038/s41586-020-2649-2}, \href {https://ui.adsabs.harvard.edu/abs/2020Natur.585..357H} {585, 357}

\bibitem[\protect\citeauthoryear{Hawley, Gammie  \& Balbus}{Hawley et~al.}{1995}]{1995ApJ...440..742H}
Hawley J.~F.,  Gammie C.~F.,   Balbus S.~A.,  1995, \mn@doi [\apj] {10.1086/175311}, 440, 742

\bibitem[\protect\citeauthoryear{{Hendler}, {Pascucci}, {Pinilla}, {Tazzari}, {Carpenter}, {Malhotra}  \& {Testi}}{{Hendler} et~al.}{2020}]{2020ApJ...895..126H}
{Hendler} N.,  {Pascucci} I.,  {Pinilla} P.,  {Tazzari} M.,  {Carpenter} J.,  {Malhotra} R.,   {Testi} L.,  2020, \mn@doi [\apj] {10.3847/1538-4357/ab70ba}, \href {https://ui.adsabs.harvard.edu/abs/2020ApJ...895..126H} {895, 126}

\bibitem[\protect\citeauthoryear{{Henning} \& {Stognienko}}{{Henning} \& {Stognienko}}{1996}]{1996A&A...311..291H}
{Henning} T.,  {Stognienko} R.,  1996, \aap, \href {https://ui.adsabs.harvard.edu/abs/1996A&A...311..291H} {311, 291}

\bibitem[\protect\citeauthoryear{Huang et~al.,}{Huang et~al.}{2018}]{2018ApJ...869L..42H}
Huang J.,  et~al., 2018, \mn@doi [\apjl] {10.3847/2041-8213/aaf740}, 869, L42

\bibitem[\protect\citeauthoryear{{Huang}, {Ansdell}, {Birnstiel}, {Czekala}, {Long}, {Williams}, {Zhang}  \& {Zhu}}{{Huang} et~al.}{2024}]{2024arXiv241003823H}
{Huang} J.,  {Ansdell} M.,  {Birnstiel} T.,  {Czekala} I.,  {Long} F.,  {Williams} J.,  {Zhang} S.,   {Zhu} Z.,  2024, \mn@doi [arXiv e-prints] {10.48550/arXiv.2410.03823}, \href {https://ui.adsabs.harvard.edu/abs/2024arXiv241003823H} {p. arXiv:2410.03823}

\bibitem[\protect\citeauthoryear{Hunter}{Hunter}{2007}]{matplotlib}
Hunter J.~D.,  2007, \mn@doi [Computing in Science & Engineering] {10.1109/MCSE.2007.55}, 9, 90

\bibitem[\protect\citeauthoryear{{Ilee}, {Walsh}, {Jennings}, {Booth}, {Rosotti}, {Teague}, {Tsukagoshi}  \& {Nomura}}{{Ilee} et~al.}{2022}]{2022MNRAS.515L..23I}
{Ilee} J.~D.,  {Walsh} C.,  {Jennings} J.,  {Booth} R.~A.,  {Rosotti} G.~P.,  {Teague} R.,  {Tsukagoshi} T.,   {Nomura} H.,  2022, \mn@doi [\mnras] {10.1093/mnrasl/slac048}, \href {https://ui.adsabs.harvard.edu/abs/2022MNRAS.515L..23I} {515, L23}

\bibitem[\protect\citeauthoryear{{Jennings}, {Tazzari}, {Clarke}, {Booth}  \& {Rosotti}}{{Jennings} et~al.}{2022}]{2022MNRAS.514.6053J}
{Jennings} J.,  {Tazzari} M.,  {Clarke} C.~J.,  {Booth} R.~A.,   {Rosotti} G.~P.,  2022, \mn@doi [\mnras] {10.1093/mnras/stac1770}, \href {https://ui.adsabs.harvard.edu/abs/2022MNRAS.514.6053J} {514, 6053}

\bibitem[\protect\citeauthoryear{{Jiang}, {Mac{\'\i}as}, {Guerra-Alvarado}  \& {Carrasco-Gonz{\'a}lez}}{{Jiang} et~al.}{2024}]{2024A&A...682A..32J}
{Jiang} H.,  {Mac{\'\i}as} E.,  {Guerra-Alvarado} O.~M.,   {Carrasco-Gonz{\'a}lez} C.,  2024, \mn@doi [\aap] {10.1051/0004-6361/202348271}, \href {https://ui.adsabs.harvard.edu/abs/2024A&A...682A..32J} {682, A32}

\bibitem[\protect\citeauthoryear{{Kanagawa}, {Ueda}, {Muto}  \& {Okuzumi}}{{Kanagawa} et~al.}{2017}]{2017ApJ...844..142K}
{Kanagawa} K.~D.,  {Ueda} T.,  {Muto} T.,   {Okuzumi} S.,  2017, \mn@doi [\apj] {10.3847/1538-4357/aa7ca1}, \href {https://ui.adsabs.harvard.edu/abs/2017ApJ...844..142K} {844, 142}

\bibitem[\protect\citeauthoryear{{Kanagawa}, {Tanaka}  \& {Szuszkiewicz}}{{Kanagawa} et~al.}{2018}]{2018ApJ...861..140K}
{Kanagawa} K.~D.,  {Tanaka} H.,   {Szuszkiewicz} E.,  2018, \mn@doi [\apj] {10.3847/1538-4357/aac8d9}, \href {https://ui.adsabs.harvard.edu/abs/2018ApJ...861..140K} {861, 140}

\bibitem[\protect\citeauthoryear{{Kley} \& {Nelson}}{{Kley} \& {Nelson}}{2012}]{2012ARA&A..50..211K}
{Kley} W.,  {Nelson} R.~P.,  2012, \mn@doi [\araa] {10.1146/annurev-astro-081811-125523}, \href {https://ui.adsabs.harvard.edu/abs/2012ARA&A..50..211K} {50, 211}

\bibitem[\protect\citeauthoryear{Kluyver et~al.,}{Kluyver et~al.}{2016}]{jupyter}
Kluyver T.,  et~al., 2016, in Loizides F.,  Scmidt B.,  eds, Positioning and Power in Academic Publishing: Players, Agents and Agendas. IOS Press, pp 87--90, \url {https://eprints.soton.ac.uk/403913/}

\bibitem[\protect\citeauthoryear{{Krijt}, {Ormel}, {Dominik}  \& {Tielens}}{{Krijt} et~al.}{2015}]{2015A&A...574A..83K}
{Krijt} S.,  {Ormel} C.~W.,  {Dominik} C.,   {Tielens} A.~G.~G.~M.,  2015, \mn@doi [\aap] {10.1051/0004-6361/201425222}, \href {https://ui.adsabs.harvard.edu/abs/2015A&A...574A..83K} {574, A83}

\bibitem[\protect\citeauthoryear{{Kuffmeier}, {Jensen}  \& {Haugb{\o}lle}}{{Kuffmeier} et~al.}{2023}]{2023EPJP..138..272K}
{Kuffmeier} M.,  {Jensen} S.~S.,   {Haugb{\o}lle} T.,  2023, \mn@doi [European Physical Journal Plus] {10.1140/epjp/s13360-023-03880-y}, \href {https://ui.adsabs.harvard.edu/abs/2023EPJP..138..272K} {138, 272}

\bibitem[\protect\citeauthoryear{Kurtovic et~al.,}{Kurtovic et~al.}{2021}]{2021A&A...645A.139K}
Kurtovic N.,  et~al., 2021, \mn@doi [\aap] {10.1051/0004-6361/202038983}, 645, A139

\bibitem[\protect\citeauthoryear{Kuznetsova, Bae, Hartmann  \& Mac~Low}{Kuznetsova et~al.}{2022}]{kuznetsova2022anisotropic}
Kuznetsova A.,  Bae J.,  Hartmann L.,   Mac~Low M.-M.,  2022, The Astrophysical Journal, 928, 92

\bibitem[\protect\citeauthoryear{{Lee}, {Tabone}, {Cabrit}, {Codella}, {Podio}, {Ferreira}  \& {Jacquemin-Ide}}{{Lee} et~al.}{2021}]{2021ApJ...907L..41L}
{Lee} C.-F.,  {Tabone} B.,  {Cabrit} S.,  {Codella} C.,  {Podio} L.,  {Ferreira} J.,   {Jacquemin-Ide} J.,  2021, \mn@doi [\apjl] {10.3847/2041-8213/abda38}, \href {https://ui.adsabs.harvard.edu/abs/2021ApJ...907L..41L} {907, L41}

\bibitem[\protect\citeauthoryear{{Lesur} et~al.,}{{Lesur} et~al.}{2023}]{2023ASPC..534..465L}
{Lesur} G.,  et~al., 2023, in {Inutsuka} S.,  {Aikawa} Y.,  {Muto} T.,  {Tomida} K.,   {Tamura} M.,  eds,  Astronomical Society of the Pacific Conference Series Vol. 534, Protostars and Planets VII. p.~465 (\mn@eprint {arXiv} {2203.09821}), \mn@doi{10.48550/arXiv.2203.09821}

\bibitem[\protect\citeauthoryear{{Li}, {Finn}, {Lovelace}  \& {Colgate}}{{Li} et~al.}{2000}]{2000ApJ...533.1023L}
{Li} H.,  {Finn} J.~M.,  {Lovelace} R.~V.~E.,   {Colgate} S.~A.,  2000, \mn@doi [\apj] {10.1086/308693}, \href {https://ui.adsabs.harvard.edu/abs/2000ApJ...533.1023L} {533, 1023}

\bibitem[\protect\citeauthoryear{{Long} et~al.,}{{Long} et~al.}{2019}]{2019ApJ...882...49L}
{Long} F.,  et~al., 2019, \mn@doi [\apj] {10.3847/1538-4357/ab2d2d}, \href {https://ui.adsabs.harvard.edu/abs/2019ApJ...882...49L} {882, 49}

\bibitem[\protect\citeauthoryear{Louvet, Dougados, Cabrit, Mardones, M{\'{e}}nard, Tabone, Pinte  \& Dent}{Louvet et~al.}{2018}]{2018A&A...618A.120L}
Louvet F.,  Dougados C.,  Cabrit S.,  Mardones D.,  M{\'{e}}nard F.,  Tabone B.,  Pinte C.,   Dent W.,  2018, \mn@doi [\aap] {10.1051/0004-6361/201731733}, 618, A120

\bibitem[\protect\citeauthoryear{{Lynden-Bell} \& {Pringle}}{{Lynden-Bell} \& {Pringle}}{1974}]{1974MNRAS.168..603L}
{Lynden-Bell} D.,  {Pringle} J.~E.,  1974, \mn@doi [\mnras] {10.1093/mnras/168.3.603}, \href {https://ui.adsabs.harvard.edu/abs/1974MNRAS.168..603L} {168, 603}

\bibitem[\protect\citeauthoryear{{Lyra}, {Johansen}, {Zsom}, {Klahr}  \& {Piskunov}}{{Lyra} et~al.}{2009}]{2009A&A...497..869L}
{Lyra} W.,  {Johansen} A.,  {Zsom} A.,  {Klahr} H.,   {Piskunov} N.,  2009, \mn@doi [\aap] {10.1051/0004-6361/200811265}, \href {https://ui.adsabs.harvard.edu/abs/2009A&A...497..869L} {497, 869}

\bibitem[\protect\citeauthoryear{{Manara} et~al.,}{{Manara} et~al.}{2019}]{2019A&A...628A..95M}
{Manara} C.~F.,  et~al., 2019, \mn@doi [\aap] {10.1051/0004-6361/201935964}, \href {https://ui.adsabs.harvard.edu/abs/2019A&A...628A..95M} {628, A95}

\bibitem[\protect\citeauthoryear{{Mathis}, {Rumpl}  \& {Nordsieck}}{{Mathis} et~al.}{1977}]{1977ApJ...217..425M}
{Mathis} J.~S.,  {Rumpl} W.,   {Nordsieck} K.~H.,  1977, \mn@doi [\apj] {10.1086/155591}, \href {https://ui.adsabs.harvard.edu/abs/1977ApJ...217..425M} {217, 425}

\bibitem[\protect\citeauthoryear{{Maureira} et~al.,}{{Maureira} et~al.}{2024}]{2024A&A...689L...5M}
{Maureira} M.~J.,  et~al., 2024, \mn@doi [\aap] {10.1051/0004-6361/202451166}, \href {https://ui.adsabs.harvard.edu/abs/2024A&A...689L...5M} {689, L5}

\bibitem[\protect\citeauthoryear{{Michoulier}, {Gonzalez}  \& {Price}}{{Michoulier} et~al.}{2024}]{2024A&A...688A..31M}
{Michoulier} S.,  {Gonzalez} J.-F.,   {Price} D.~J.,  2024, \mn@doi [\aap] {10.1051/0004-6361/202449719}, \href {https://ui.adsabs.harvard.edu/abs/2024A&A...688A..31M} {688, A31}

\bibitem[\protect\citeauthoryear{{Miley} et~al.,}{{Miley} et~al.}{2024}]{2024A&A...682A..55M}
{Miley} J.~M.,  et~al., 2024, \mn@doi [\aap] {10.1051/0004-6361/202347135}, \href {https://ui.adsabs.harvard.edu/abs/2024A&A...682A..55M} {682, A55}

\bibitem[\protect\citeauthoryear{{Miotello}, {Rosotti}, {Ansdell}, {Facchini}, {Manara}, {Williams}  \& {Bruderer}}{{Miotello} et~al.}{2021}]{2021A&A...651A..48M}
{Miotello} A.,  {Rosotti} G.,  {Ansdell} M.,  {Facchini} S.,  {Manara} C.~F.,  {Williams} J.~P.,   {Bruderer} S.,  2021, \mn@doi [\aap] {10.1051/0004-6361/202140550}, \href {https://ui.adsabs.harvard.edu/abs/2021A&A...651A..48M} {651, A48}

\bibitem[\protect\citeauthoryear{{Morishima}}{{Morishima}}{2012}]{2012MNRAS.420.2851M}
{Morishima} R.,  2012, \mn@doi [\mnras] {10.1111/j.1365-2966.2011.19940.x}, \href {https://ui.adsabs.harvard.edu/abs/2012MNRAS.420.2851M} {420, 2851}

\bibitem[\protect\citeauthoryear{{Musiolik} \& {Wurm}}{{Musiolik} \& {Wurm}}{2019}]{2019ApJ...873...58M}
{Musiolik} G.,  {Wurm} G.,  2019, \mn@doi [\apj] {10.3847/1538-4357/ab0428}, \href {https://ui.adsabs.harvard.edu/abs/2019ApJ...873...58M} {873, 58}

\bibitem[\protect\citeauthoryear{{Nakagawa}, {Sekiya}  \& {Hayashi}}{{Nakagawa} et~al.}{1986}]{1986Icar...67..375N}
{Nakagawa} Y.,  {Sekiya} M.,   {Hayashi} C.,  1986, \mn@doi [\icarus] {10.1016/0019-1035(86)90121-1}, \href {https://ui.adsabs.harvard.edu/abs/1986Icar...67..375N} {67, 375}

\bibitem[\protect\citeauthoryear{{Nazari}, {Sellek}  \& {Rosotti}}{{Nazari} et~al.}{2024}]{2024arXiv241009042N}
{Nazari} P.,  {Sellek} A.~D.,   {Rosotti} G.~P.,  2024, \mn@doi [arXiv e-prints] {10.48550/arXiv.2410.09042}, \href {https://ui.adsabs.harvard.edu/abs/2024arXiv241009042N} {p. arXiv:2410.09042}

\bibitem[\protect\citeauthoryear{Nelson, Gressel  \& Umurhan}{Nelson et~al.}{2013}]{2013MNRAS.435.2610N}
Nelson R.~P.,  Gressel O.,   Umurhan O.~M.,  2013, \mn@doi [\mnras] {10.1093/mnras/stt1475}, 435, 2610

\bibitem[\protect\citeauthoryear{{Ohashi} \& {Kataoka}}{{Ohashi} \& {Kataoka}}{2019}]{2019ApJ...886..103O}
{Ohashi} S.,  {Kataoka} A.,  2019, \mn@doi [\apj] {10.3847/1538-4357/ab5107}, \href {https://ui.adsabs.harvard.edu/abs/2019ApJ...886..103O} {886, 103}

\bibitem[\protect\citeauthoryear{{Okuzumi}, {Momose}, {Sirono}, {Kobayashi}  \& {Tanaka}}{{Okuzumi} et~al.}{2016}]{2016ApJ...821...82O}
{Okuzumi} S.,  {Momose} M.,  {Sirono} S.-i.,  {Kobayashi} H.,   {Tanaka} H.,  2016, \mn@doi [\apj] {10.3847/0004-637X/821/2/82}, \href {https://ui.adsabs.harvard.edu/abs/2016ApJ...821...82O} {821, 82}

\bibitem[\protect\citeauthoryear{{Ormel} \& {Cuzzi}}{{Ormel} \& {Cuzzi}}{2007}]{2007A&A...466..413O}
{Ormel} C.~W.,  {Cuzzi} J.~N.,  2007, \mn@doi [\aap] {10.1051/0004-6361:20066899}, \href {https://ui.adsabs.harvard.edu/abs/2007A&A...466..413O} {466, 413}

\bibitem[\protect\citeauthoryear{Otter, Ginsburg, Ballering, Bally, Eisner, Goddi, Plambeck  \& Wright}{Otter et~al.}{2021}]{2021ApJ...923..221O}
Otter J.,  Ginsburg A.,  Ballering N.~P.,  Bally J.,  Eisner J.,  Goddi C.,  Plambeck R.,   Wright M.,  2021, \mn@doi [\apj] {10.3847/1538-4357/ac29c2}, 923, 221

\bibitem[\protect\citeauthoryear{Pascucci et~al.,}{Pascucci et~al.}{2016}]{2016ApJ...831..125P}
Pascucci I.,  et~al., 2016, \mn@doi [\apj] {10.3847/0004-637X/831/2/125}, 831, 125

\bibitem[\protect\citeauthoryear{{Pecaut}, {Mamajek}  \& {Bubar}}{{Pecaut} et~al.}{2012}]{2012ApJ...746..154P}
{Pecaut} M.~J.,  {Mamajek} E.~E.,   {Bubar} E.~J.,  2012, \mn@doi [\apj] {10.1088/0004-637X/746/2/154}, \href {https://ui.adsabs.harvard.edu/abs/2012ApJ...746..154P} {746, 154}

\bibitem[\protect\citeauthoryear{{Pi{\'e}tu}, {Guilloteau}, {Di Folco}, {Dutrey}  \& {Boehler}}{{Pi{\'e}tu} et~al.}{2014}]{2014A&A...564A..95P}
{Pi{\'e}tu} V.,  {Guilloteau} S.,  {Di Folco} E.,  {Dutrey} A.,   {Boehler} Y.,  2014, \mn@doi [\aap] {10.1051/0004-6361/201322388}, \href {https://ui.adsabs.harvard.edu/abs/2014A&A...564A..95P} {564, A95}

\bibitem[\protect\citeauthoryear{{Pinilla}, {Birnstiel}, {Ricci}, {Dullemond}, {Uribe}, {Testi}  \& {Natta}}{{Pinilla} et~al.}{2012}]{2012A&A...538A...114P}
{Pinilla} P.,  {Birnstiel} T.,  {Ricci} L.,  {Dullemond} C.~P.,  {Uribe} A.~L.,  {Testi} L.,   {Natta} A.,  2012, \mn@doi [\aap] {10.1051/0004-6361/201118204}, \href {https://ui.adsabs.harvard.edu/abs/2012A&A...538A.114P} {538, A114}

\bibitem[\protect\citeauthoryear{{Pinilla}, {Flock}, {Ovelar}  \& {Birnstiel}}{{Pinilla} et~al.}{2016}]{2016A&A...596A..81P}
{Pinilla} P.,  {Flock} M.,  {Ovelar} M. d.~J.,   {Birnstiel} T.,  2016, \mn@doi [\aap] {10.1051/0004-6361/201628441}, \href {https://ui.adsabs.harvard.edu/abs/2016A&A...596A..81P} {596, A81}

\bibitem[\protect\citeauthoryear{{Pinilla} et~al.,}{{Pinilla} et~al.}{2017a}]{2017ApJ...839...99P}
{Pinilla} P.,  et~al., 2017a, \mn@doi [\apj] {10.3847/1538-4357/aa6973}, \href {https://ui.adsabs.harvard.edu/abs/2017ApJ...839...99P} {839, 99}

\bibitem[\protect\citeauthoryear{{Pinilla}, {Pohl}, {Stammler}  \& {Birnstiel}}{{Pinilla} et~al.}{2017b}]{2017ApJ...845...68P}
{Pinilla} P.,  {Pohl} A.,  {Stammler} S.~M.,   {Birnstiel} T.,  2017b, \mn@doi [\apj] {10.3847/1538-4357/aa7edb}, \href {https://ui.adsabs.harvard.edu/abs/2017ApJ...845...68P} {845, 68}

\bibitem[\protect\citeauthoryear{{Pinilla} et~al.,}{{Pinilla} et~al.}{2018}]{2018ApJ...859...32P}
{Pinilla} P.,  et~al., 2018, \mn@doi [\apj] {10.3847/1538-4357/aabf94}, \href {https://ui.adsabs.harvard.edu/abs/2018ApJ...859...32P} {859, 32}

\bibitem[\protect\citeauthoryear{{Pinilla}, {Benisty}, {Cazzoletti}, {Harsono}, {P{\'e}rez}  \& {Tazzari}}{{Pinilla} et~al.}{2019}]{2019ApJ...878...16P}
{Pinilla} P.,  {Benisty} M.,  {Cazzoletti} P.,  {Harsono} D.,  {P{\'e}rez} L.~M.,   {Tazzari} M.,  2019, \mn@doi [\apj] {10.3847/1538-4357/ab1cb8}, \href {https://ui.adsabs.harvard.edu/abs/2019ApJ...878...16P} {878, 16}

\bibitem[\protect\citeauthoryear{{Pinilla}, {Lenz}  \& {Stammler}}{{Pinilla} et~al.}{2021a}]{2021A&A...645A..70P}
{Pinilla} P.,  {Lenz} C.~T.,   {Stammler} S.~M.,  2021a, \mn@doi [\aap] {10.1051/0004-6361/202038920}, \href {https://ui.adsabs.harvard.edu/abs/2021A&A...645A..70P} {645, A70}

\bibitem[\protect\citeauthoryear{{Pinilla} et~al.,}{{Pinilla} et~al.}{2021b}]{2021A&A...649A.122P}
{Pinilla} P.,  et~al., 2021b, \mn@doi [\aap] {10.1051/0004-6361/202140371}, \href {https://ui.adsabs.harvard.edu/abs/2021A&A...649A.122P} {649, A122}

\bibitem[\protect\citeauthoryear{{Preibisch}, {Brown}, {Bridges}, {Guenther}  \& {Zinnecker}}{{Preibisch} et~al.}{2002}]{2002AJ....124..404P}
{Preibisch} T.,  {Brown} A. G.~A.,  {Bridges} T.,  {Guenther} E.,   {Zinnecker} H.,  2002, \mn@doi [\aj] {10.1086/341174}, \href {https://ui.adsabs.harvard.edu/abs/2002AJ....124..404P} {124, 404}

\bibitem[\protect\citeauthoryear{{Ribas} et~al.,}{{Ribas} et~al.}{2023}]{2023A&A...673A..77R}
{Ribas} {\'A}.,  et~al., 2023, \mn@doi [\aap] {10.1051/0004-6361/202245637}, \href {https://ui.adsabs.harvard.edu/abs/2023A&A...673A..77R} {673, A77}

\bibitem[\protect\citeauthoryear{{Rodenkirch} \& {Dullemond}}{{Rodenkirch} \& {Dullemond}}{2022}]{2022A&A...659A..42R}
{Rodenkirch} P.~J.,  {Dullemond} C.~P.,  2022, \mn@doi [\aap] {10.1051/0004-6361/202142571}, \href {https://ui.adsabs.harvard.edu/abs/2022A&A...659A..42R} {659, A42}

\bibitem[\protect\citeauthoryear{{Rosotti}}{{Rosotti}}{2023}]{2023NewAR..9601674R}
{Rosotti} G.~P.,  2023, \mn@doi [\nar] {10.1016/j.newar.2023.101674}, \href {https://ui.adsabs.harvard.edu/abs/2023NewAR..9601674R} {96, 101674}

\bibitem[\protect\citeauthoryear{Rosotti, Tazzari, Booth, Testi, Lodato  \& Clarke}{Rosotti et~al.}{2019}]{2019MNRAS.486.4829R}
Rosotti G.~P.,  Tazzari M.,  Booth R.~A.,  Testi L.,  Lodato G.,   Clarke C.,  2019, \mn@doi [\mnras] {10.1093/mnras/stz1190}, 486, 4829

\bibitem[\protect\citeauthoryear{{Rota} et~al.,}{{Rota} et~al.}{2022}]{2022A&A...662A.121R}
{Rota} A.~A.,  et~al., 2022, \mn@doi [\aap] {10.1051/0004-6361/202141035}, \href {https://ui.adsabs.harvard.edu/abs/2022A&A...662A.121R} {662, A121}

\bibitem[\protect\citeauthoryear{{Salmeron} \& {Wardle}}{{Salmeron} \& {Wardle}}{2008}]{2008MNRAS.388.1223S}
{Salmeron} R.,  {Wardle} M.,  2008, \mn@doi [\mnras] {10.1111/j.1365-2966.2008.13430.x}, \href {https://ui.adsabs.harvard.edu/abs/2008MNRAS.388.1223S} {388, 1223}

\bibitem[\protect\citeauthoryear{{S{\'a}ndor}, {Guilera}, {Reg{\'a}ly}  \& {Lyra}}{{S{\'a}ndor} et~al.}{2024}]{2024A&A...686A..78S}
{S{\'a}ndor} Z.,  {Guilera} O.~M.,  {Reg{\'a}ly} Z.,   {Lyra} W.,  2024, \mn@doi [\aap] {10.1051/0004-6361/202347605}, \href {https://ui.adsabs.harvard.edu/abs/2024A&A...686A..78S} {686, A78}

\bibitem[\protect\citeauthoryear{{Sano}, {Miyama}, {Umebayashi}  \& {Nakano}}{{Sano} et~al.}{2000}]{2000ApJ...543..486S}
{Sano} T.,  {Miyama} S.~M.,  {Umebayashi} T.,   {Nakano} T.,  2000, \mn@doi [\apj] {10.1086/317075}, \href {https://ui.adsabs.harvard.edu/abs/2000ApJ...543..486S} {543, 486}

\bibitem[\protect\citeauthoryear{{Segura-Cox} et~al.,}{{Segura-Cox} et~al.}{2020}]{2020Natur.586..228S}
{Segura-Cox} D.~M.,  et~al., 2020, \mn@doi [\nat] {10.1038/s41586-020-2779-6}, \href {https://ui.adsabs.harvard.edu/abs/2020Natur.586..228S} {586, 228}

\bibitem[\protect\citeauthoryear{{Sellek}, {Vlasblom}  \& {van Dishoeck}}{{Sellek} et~al.}{2024}]{2024arXiv241201895S}
{Sellek} A.~D.,  {Vlasblom} M.,   {van Dishoeck} E.~F.,  2024, \mn@doi [arXiv e-prints] {10.48550/arXiv.2412.01895}, \href {https://ui.adsabs.harvard.edu/abs/2024arXiv241201895S} {p. arXiv:2412.01895}

\bibitem[\protect\citeauthoryear{{Shakura} \& {Sunyaev}}{{Shakura} \& {Sunyaev}}{1973}]{1973A&A....24..337S}
{Shakura} N.~I.,  {Sunyaev} R.~A.,  1973, \aap, \href {https://ui.adsabs.harvard.edu/abs/1973A&A....24..337S} {24, 337}

\bibitem[\protect\citeauthoryear{{Shi} et~al.,}{{Shi} et~al.}{2024}]{2024ApJ...966...59S}
{Shi} Y.,  et~al., 2024, \mn@doi [\apj] {10.3847/1538-4357/ad2e94}, \href {https://ui.adsabs.harvard.edu/abs/2024ApJ...966...59S} {966, 59}

\bibitem[\protect\citeauthoryear{{Smoluchowski}}{{Smoluchowski}}{1916}]{1916ZPhy...17..557S}
{Smoluchowski} M.~V.,  1916, Zeitschrift fur Physik, \href {https://ui.adsabs.harvard.edu/abs/1916ZPhy...17..557S} {17, 557}

\bibitem[\protect\citeauthoryear{{Speedie} et~al.,}{{Speedie} et~al.}{2024}]{2024Natur.633...58S}
{Speedie} J.,  et~al., 2024, \mn@doi [\nat] {10.1038/s41586-024-07877-0}, \href {https://ui.adsabs.harvard.edu/abs/2024Natur.633...58S} {633, 58}

\bibitem[\protect\citeauthoryear{{Stadler}, {G{\'a}rate}, {Pinilla}, {Lenz}, {Dullemond}, {Birnstiel}  \& {Stammler}}{{Stadler} et~al.}{2022}]{2022A&A...668A.104S}
{Stadler} J.,  {G{\'a}rate} M.,  {Pinilla} P.,  {Lenz} C.,  {Dullemond} C.~P.,  {Birnstiel} T.,   {Stammler} S.~M.,  2022, \mn@doi [\aap] {10.1051/0004-6361/202243338}, \href {https://ui.adsabs.harvard.edu/abs/2022A&A...668A.104S} {668, A104}

\bibitem[\protect\citeauthoryear{{Stammler} \& {Birnstiel}}{{Stammler} \& {Birnstiel}}{2022}]{2022ApJ...935...35S}
{Stammler} S.~M.,  {Birnstiel} T.,  2022, \mn@doi [\apj] {10.3847/1538-4357/ac7d58}, \href {https://ui.adsabs.harvard.edu/abs/2022ApJ...935...35S} {935, 35}

\bibitem[\protect\citeauthoryear{{Stammler}, {Birnstiel}  \& {G{\'a}rate}}{{Stammler} et~al.}{2023a}]{2023ascl.soft10005S}
{Stammler} S.~M.,  {Birnstiel} T.,   {G{\'a}rate} M.,  2023a, {DustPyLib: A library of DustPy extensions}, Astrophysics Source Code Library, record ascl:2310.005

\bibitem[\protect\citeauthoryear{{Stammler}, {Lichtenberg}, {Dr{\k{a}}{\.z}kowska}  \& {Birnstiel}}{{Stammler} et~al.}{2023b}]{2023A&A...670L...5S}
{Stammler} S.~M.,  {Lichtenberg} T.,  {Dr{\k{a}}{\.z}kowska} J.,   {Birnstiel} T.,  2023b, \mn@doi [\aap] {10.1051/0004-6361/202245512}, \href {https://ui.adsabs.harvard.edu/abs/2023A&A...670L...5S} {670, L5}

\bibitem[\protect\citeauthoryear{Suzuki, Ogihara, Morbidelli, Crida  \& Guillot}{Suzuki et~al.}{2016}]{2016A&A...596A..74S}
Suzuki T.~K.,  Ogihara M.,  Morbidelli A.,  Crida A.,   Guillot T.,  2016, \mn@doi [\aap] {10.1051/0004-6361/201628955}, 596, A74

\bibitem[\protect\citeauthoryear{{Tabone} et~al.,}{{Tabone} et~al.}{2017}]{2017A&A...607L...6T}
{Tabone} B.,  et~al., 2017, \mn@doi [\aap] {10.1051/0004-6361/201731691}, 607, L6

\bibitem[\protect\citeauthoryear{{Tabone}, {Rosotti}, {Cridland}, {Armitage}  \& {Lodato}}{{Tabone} et~al.}{2022}]{2022MNRAS.512.2290T}
{Tabone} B.,  {Rosotti} G.~P.,  {Cridland} A.~J.,  {Armitage} P.~J.,   {Lodato} G.,  2022, \mn@doi [\mnras] {10.1093/mnras/stab3442}, \href {https://ui.adsabs.harvard.edu/abs/2022MNRAS.512.2290T} {512, 2290}

\bibitem[\protect\citeauthoryear{{Takeuchi} \& {Lin}}{{Takeuchi} \& {Lin}}{2002}]{2002ApJ...581.1344T}
{Takeuchi} T.,  {Lin} D.~N.~C.,  2002, \mn@doi [\apj] {10.1086/344437}, \href {https://ui.adsabs.harvard.edu/abs/2002ApJ...581.1344T} {581, 1344}

\bibitem[\protect\citeauthoryear{{Tazaki}, {Tanaka}, {Kataoka}, {Okuzumi}  \& {Muto}}{{Tazaki} et~al.}{2019}]{2019ApJ...885...52T}
{Tazaki} R.,  {Tanaka} H.,  {Kataoka} A.,  {Okuzumi} S.,   {Muto} T.,  2019, \mn@doi [\apj] {10.3847/1538-4357/ab45f0}, \href {https://ui.adsabs.harvard.edu/abs/2019ApJ...885...52T} {885, 52}

\bibitem[\protect\citeauthoryear{{Tazaki}, {Ginski}  \& {Dominik}}{{Tazaki} et~al.}{2023}]{2023ApJ...944L..43T}
{Tazaki} R.,  {Ginski} C.,   {Dominik} C.,  2023, \mn@doi [\apjl] {10.3847/2041-8213/acb824}, \href {https://ui.adsabs.harvard.edu/abs/2023ApJ...944L..43T} {944, L43}

\bibitem[\protect\citeauthoryear{{Tazzari} et~al.,}{{Tazzari} et~al.}{2017}]{2017A&A...606A..88T}
{Tazzari} M.,  et~al., 2017, \mn@doi [\aap] {10.1051/0004-6361/201730890}, \href {https://ui.adsabs.harvard.edu/abs/2017A&A...606A..88T} {606, A88}

\bibitem[\protect\citeauthoryear{{Tazzari}, {Clarke}, {Testi}, {Williams}, {Facchini}, {Manara}, {Natta}  \& {Rosotti}}{{Tazzari} et~al.}{2021}]{2021MNRAS.506.2804T}
{Tazzari} M.,  {Clarke} C.~J.,  {Testi} L.,  {Williams} J.~P.,  {Facchini} S.,  {Manara} C.~F.,  {Natta} A.,   {Rosotti} G.,  2021, \mn@doi [\mnras] {10.1093/mnras/stab1808}, \href {https://ui.adsabs.harvard.edu/abs/2021MNRAS.506.2804T} {506, 2804}

\bibitem[\protect\citeauthoryear{{The Astropy Collaboration} et~al.,}{{The Astropy Collaboration} et~al.}{2018}]{astropy2018}
{The Astropy Collaboration} et~al., 2018, \mn@doi [\aj] {10.3847/1538-3881/aabc4f}, \href {https://ui.adsabs.harvard.edu/abs/2018AJ....156..123T} {156, 123}

\bibitem[\protect\citeauthoryear{{Tong}, {Alexander}  \& {Rosotti}}{{Tong} et~al.}{2024}]{2024_disc_evo}
{Tong} S.,  {Alexander} R.,   {Rosotti} G.,  2024, \mn@doi [\mnras] {10.1006/icar.1995.1058}, \href {https://ui.adsabs.harvard.edu/abs/1995Icar..114..237D} {114, 237}

\bibitem[\protect\citeauthoryear{{Trapman}, {Ansdell}, {Hogerheijde}, {Facchini}, {Manara}, {Miotello}, {Williams}  \& {Bruderer}}{{Trapman} et~al.}{2020}]{2020A&A...638A..38T}
{Trapman} L.,  {Ansdell} M.,  {Hogerheijde} M.~R.,  {Facchini} S.,  {Manara} C.~F.,  {Miotello} A.,  {Williams} J.~P.,   {Bruderer} S.,  2020, \mn@doi [\aap] {10.1051/0004-6361/201834537}, \href {https://ui.adsabs.harvard.edu/abs/2020A&A...638A..38T} {638, A38}

\bibitem[\protect\citeauthoryear{{Tripathi}, {Andrews}, {Birnstiel}  \& {Wilner}}{{Tripathi} et~al.}{2017}]{2017ApJ...845...44T}
{Tripathi} A.,  {Andrews} S.~M.,  {Birnstiel} T.,   {Wilner} D.~J.,  2017, \mn@doi [\apj] {10.3847/1538-4357/aa7c62}, \href {https://ui.adsabs.harvard.edu/abs/2017ApJ...845...44T} {845, 44}

\bibitem[\protect\citeauthoryear{{Ueda}, {Kataoka}  \& {Tsukagoshi}}{{Ueda} et~al.}{2022}]{2022ApJ...930...56U}
{Ueda} T.,  {Kataoka} A.,   {Tsukagoshi} T.,  2022, \mn@doi [\apj] {10.3847/1538-4357/ac634d}, \href {https://ui.adsabs.harvard.edu/abs/2022ApJ...930...56U} {930, 56}

\bibitem[\protect\citeauthoryear{{Ueda}, {Tazaki}, {Okuzumi}, {Flock}  \& {Sudarshan}}{{Ueda} et~al.}{2024}]{2024NatAs...8.1148U}
{Ueda} T.,  {Tazaki} R.,  {Okuzumi} S.,  {Flock} M.,   {Sudarshan} P.,  2024, \mn@doi [Nature Astronomy] {10.1038/s41550-024-02308-6}, \href {https://ui.adsabs.harvard.edu/abs/2024NatAs...8.1148U} {8, 1148}

\bibitem[\protect\citeauthoryear{{Villenave} et~al.,}{{Villenave} et~al.}{2021}]{2021A&A...653A..46V}
{Villenave} M.,  et~al., 2021, \mn@doi [\aap] {10.1051/0004-6361/202140496}, \href {https://ui.adsabs.harvard.edu/abs/2021A&A...653A..46V} {653, A46}

\bibitem[\protect\citeauthoryear{Virtanen et~al.,}{Virtanen et~al.}{2020}]{scipy2020}
Virtanen P.,  et~al., 2020, \mn@doi [Nature Methods] {10.1038/s41592-019-0686-2}, 17, 261

\bibitem[\protect\citeauthoryear{Vlasblom et~al.,}{Vlasblom et~al.}{2024}]{vlasblom2024}
Vlasblom M.,  et~al., 2024, MINDS. JWST-MIRI reveals a peculiar CO$_2$-rich chemistry in the drift-dominated disk CX Tau (\mn@eprint {arXiv} {2412.12715}), \url {https://arxiv.org/abs/2412.12715}

\bibitem[\protect\citeauthoryear{{Warren} \& {Brandt}}{{Warren} \& {Brandt}}{2008}]{2008JGRD..11314220W}
{Warren} S.~G.,  {Brandt} R.~E.,  2008, \mn@doi [Journal of Geophysical Research (Atmospheres)] {10.1029/2007JD009744}, \href {https://ui.adsabs.harvard.edu/abs/2008JGRD..11314220W} {113, D14220}

\bibitem[\protect\citeauthoryear{{Weidenschilling}}{{Weidenschilling}}{1977}]{1977MNRAS.180...57W}
{Weidenschilling} S.~J.,  1977, \mn@doi [\mnras] {10.1093/mnras/180.2.57}, \href {https://ui.adsabs.harvard.edu/abs/1977MNRAS.180...57W} {180, 57}

\bibitem[\protect\citeauthoryear{{W}es {M}c{K}inney}{{W}es {M}c{K}inney}{2010}]{pandas2010}
{W}es {M}c{K}inney 2010, in {S}t\'efan van~der {W}alt {J}arrod {M}illman eds, {P}roceedings of the 9th {P}ython in {S}cience {C}onference. pp 56 -- 61, \mn@doi{10.25080/Majora-92bf1922-00a}

\bibitem[\protect\citeauthoryear{{Whipple}}{{Whipple}}{1972}]{1972fpp..conf..211W}
{Whipple} F.~L.,  1972, in {Elvius} A.,  ed., From Plasma to Planet. p.~211

\bibitem[\protect\citeauthoryear{{Windmark}, {Birnstiel}, {Ormel}  \& {Dullemond}}{{Windmark} et~al.}{2012}]{2012A&A...544L..16W}
{Windmark} F.,  {Birnstiel} T.,  {Ormel} C.~W.,   {Dullemond} C.~P.,  2012, \mn@doi [\aap] {10.1051/0004-6361/201220004}, \href {https://ui.adsabs.harvard.edu/abs/2012A&A...544L..16W} {544, L16}

\bibitem[\protect\citeauthoryear{{Xin}, {Espaillat}, {Rilinger}, {Ribas}  \& {Mac{\'\i}as}}{{Xin} et~al.}{2023}]{2023ApJ...942....4X}
{Xin} Z.,  {Espaillat} C.~C.,  {Rilinger} A.~M.,  {Ribas} {\'A}.,   {Mac{\'\i}as} E.,  2023, \mn@doi [\apj] {10.3847/1538-4357/aca52b}, \href {https://ui.adsabs.harvard.edu/abs/2023ApJ...942....4X} {942, 4}

\bibitem[\protect\citeauthoryear{{Yamaguchi}, {Tsukagoshi}, {Muto}, {Nomura}, {Nakazato}, {Ikeda}, {Tamura}  \& {Kawabe}}{{Yamaguchi} et~al.}{2021}]{2021ApJ...923..121Y}
{Yamaguchi} M.,  {Tsukagoshi} T.,  {Muto} T.,  {Nomura} H.,  {Nakazato} T.,  {Ikeda} S.,  {Tamura} M.,   {Kawabe} R.,  2021, \mn@doi [\apj] {10.3847/1538-4357/ac2bfd}, \href {https://ui.adsabs.harvard.edu/abs/2021ApJ...923..121Y} {923, 121}

\bibitem[\protect\citeauthoryear{{Yamaguchi} et~al.,}{{Yamaguchi} et~al.}{2024}]{2024PASJ...76..437Y}
{Yamaguchi} M.,  et~al., 2024, \mn@doi [\pasj] {10.1093/pasj/psae022}, \href {https://ui.adsabs.harvard.edu/abs/2024PASJ...76..437Y} {76, 437}

\bibitem[\protect\citeauthoryear{{Youdin} \& {Lithwick}}{{Youdin} \& {Lithwick}}{2007}]{2007Icar..192..588Y}
{Youdin} A.~N.,  {Lithwick} Y.,  2007, \mn@doi [\icarus] {10.1016/j.icarus.2007.07.012}, \href {https://ui.adsabs.harvard.edu/abs/2007Icar..192..588Y} {192, 588}

\bibitem[\protect\citeauthoryear{{Zagaria}, {Rosotti}  \& {Lodato}}{{Zagaria} et~al.}{2021a}]{2021MNRAS.504.2235Z}
{Zagaria} F.,  {Rosotti} G.~P.,   {Lodato} G.,  2021a, \mn@doi [\mnras] {10.1093/mnras/stab985}, \href {https://ui.adsabs.harvard.edu/abs/2021MNRAS.504.2235Z} {504, 2235}

\bibitem[\protect\citeauthoryear{{Zagaria}, {Rosotti}  \& {Lodato}}{{Zagaria} et~al.}{2021b}]{2021MNRAS.507.2531Z}
{Zagaria} F.,  {Rosotti} G.~P.,   {Lodato} G.,  2021b, \mn@doi [\mnras] {10.1093/mnras/stab2024}, \href {https://ui.adsabs.harvard.edu/abs/2021MNRAS.507.2531Z} {507, 2531}

\bibitem[\protect\citeauthoryear{{Zhang}, {Blake}  \& {Bergin}}{{Zhang} et~al.}{2015}]{2015ApJ...806L...7Z}
{Zhang} K.,  {Blake} G.~A.,   {Bergin} E.~A.,  2015, \mn@doi [\apjl] {10.1088/2041-8205/806/1/L7}, \href {https://ui.adsabs.harvard.edu/abs/2015ApJ...806L...7Z} {806, L7}

\bibitem[\protect\citeauthoryear{{Zhang}, {Kalscheur}, {Long}, {Zhang}, {Long}, {Bergin}, {Zhu}  \& {Trapman}}{{Zhang} et~al.}{2023a}]{2023ApJ...952..108Z}
{Zhang} S.,  {Kalscheur} M.,  {Long} F.,  {Zhang} K.,  {Long} D.~E.,  {Bergin} E.~A.,  {Zhu} Z.,   {Trapman} L.,  2023a, \mn@doi [\apj] {10.3847/1538-4357/acd334}, \href {https://ui.adsabs.harvard.edu/abs/2023ApJ...952..108Z} {952, 108}

\bibitem[\protect\citeauthoryear{{Zhang}, {Zhu}, {Ueda}, {Kataoka}, {Sierra}, {Carrasco-Gonz{\'a}lez}  \& {Mac{\'\i}as}}{{Zhang} et~al.}{2023b}]{2023ApJ...953...96Z}
{Zhang} S.,  {Zhu} Z.,  {Ueda} T.,  {Kataoka} A.,  {Sierra} A.,  {Carrasco-Gonz{\'a}lez} C.,   {Mac{\'\i}as} E.,  2023b, \mn@doi [\apj] {10.3847/1538-4357/acdb4e}, \href {https://ui.adsabs.harvard.edu/abs/2023ApJ...953...96Z} {953, 96}

\bibitem[\protect\citeauthoryear{{Zormpas}, {Birnstiel}, {Rosotti}  \& {Andrews}}{{Zormpas} et~al.}{2022}]{2022A&A...661A..66Z}
{Zormpas} A.,  {Birnstiel} T.,  {Rosotti} G.~P.,   {Andrews} S.~M.,  2022, \mn@doi [\aap] {10.1051/0004-6361/202142046}, \href {https://ui.adsabs.harvard.edu/abs/2022A&A...661A..66Z} {661, A66}

\bibitem[\protect\citeauthoryear{{de Valon}, {Dougados}, {Cabrit}, {Louvet}, {Zapata}  \& {Mardones}}{{de Valon} et~al.}{2020}]{2020A&A...634L..12D}
{de Valon} A.,  {Dougados} C.,  {Cabrit} S.,  {Louvet} F.,  {Zapata} L.~A.,   {Mardones} D.,  2020, \mn@doi [\aap] {10.1051/0004-6361/201936950}, \href {https://ui.adsabs.harvard.edu/abs/2020A&A...634L..12D} {634, L12}

\bibitem[\protect\citeauthoryear{pandas~development team}{pandas~development team}{2020}]{pandas2020}
pandas~development team T.,  2020, pandas-dev/pandas: Pandas, \mn@doi{10.5281/zenodo.3509134}, \url {https://doi.org/10.5281/zenodo.3509134}

\bibitem[\protect\citeauthoryear{{van der Marel} \& {Mulders}}{{van der Marel} \& {Mulders}}{2021}]{2021AJ....162...28V}
{van der Marel} N.,  {Mulders} G.~D.,  2021, \mn@doi [\aj] {10.3847/1538-3881/ac0255}, \href {https://ui.adsabs.harvard.edu/abs/2021AJ....162...28V} {162, 28}

\bibitem[\protect\citeauthoryear{van~der Marel et~al.,}{van~der Marel et~al.}{2013}]{2013Sci...340.1199V}
van~der Marel N.,  et~al., 2013, \mn@doi [Science] {10.1126/science.1236770}, 340, 1199

\bibitem[\protect\citeauthoryear{{van der Marel} et~al.,}{{van der Marel} et~al.}{2022}]{2022arXiv220408225V}
{van der Marel} N.,  et~al., 2022, \mn@doi [arXiv e-prints] {10.48550/arXiv.2204.08225}, \href {https://ui.adsabs.harvard.edu/abs/2022arXiv220408225V} {p. arXiv:2204.08225}

\bibitem[\protect\citeauthoryear{van~der Walt, Colbert  \& Varoquaux}{van~der Walt et~al.}{2011}]{numpy}
van~der Walt S.,  Colbert S.~C.,   Varoquaux G.,  2011, \mn@doi [Computing in Science & Engineering] {10.1109/MCSE.2011.37}, 13, 22

\makeatother
\end{thebibliography}




\appendix
\section{Dust disc properties for Models 1-11 at $\lambda = 3~\mathrm{mm}$}

\begin{table*}
\caption{Dust disc properties at $\lambda = 3~\mathrm{mm}$ for models without (Models 1-6) and with (Models 7-11) pressure bumps at $t=1$ and $3\,\mathrm{Myr}$, calculated using \textsc{DSHARP} opacities \citep{2018ApJ...869L..45B}.}\label{tb:models3}
\begin{tabular}{ccccccccc}

\hline
\hline
(1) & (2) & (3) & (4) & (5) & (6) & (7) & (8) & (9) \\
{Model} & {$\alpha_\mathrm{MRI}$} &$R_t/R_b$ & $F_\mathrm{3mm}^\mathrm{1Myr}$ &$R_{d,\mathrm{68\%}}^\mathrm{1Myr} $& $R_{d,\mathrm{90\%}}^\mathrm{1Myr}$& 
$F_\mathrm{3mm}^\mathrm{3Myr}$ & 
$R_{d,\mathrm{68\%}}^\mathrm{3Myr} $& $R_{d,\mathrm{90\%}}^\mathrm{3Myr}$\\
  & & [au/au] & [mJy] & [au] & [au]  &[mJy] & [au] & [au] \\ 
\hline
1 & $10^{-2}$ & 30/-& 0.78 & 11.91 & 23.77 & 0.21 & 18.88 & 23.77 \\
2 & $10^{-3}$ & 30/- & 1.49 & 13.37 & 23.77 & 1.12 & 15.00 & 26.67 \\
3 &  $5\times 10^{-4}$ & 30/- & 2.11 & 13.37 & 21.19 & 1.89 & 13.37 & 21.19 \\
4 & $ 10^{-3}$ (S) & 30/- & 6.38 & 16.83 & 23.77 & 3.60 & 13.37 & 21.19 \\
5 & $ 10^{-3}$ & 20/- & 1.17 & 11.91 & 18.88 & 1.03 & 11.91 & 18.88 \\
6 &  $ 10^{-3}$ & 50/- & 3.36 & 11.91 & 29.93 & 1.14 & 23.77 & 37.68 \\
7 & $10^{-3}$ & 30/6 & 2.40 & 9.46 & 16.83 & 1.84 & 9.46 & 18.88 \\
8 & $10^{-3}$ & 30/15 & 1.95 & 21.19 & 23.77 & 1.99 & 21.19 & 23.77 \\
9 & $10^{-3}$ &30/30 & 1.07 & 11.91 & 21.19 & 0.91 & 13.37 & 21.19 \\
10 & $10^{-3}$ & 30/45 & 1.03 & 9.46 & 21.19 & 0.63 & 15.00 & 23.77 \\
11 & $10^{-3}$ & 30/80 & 1.81 & 13.37 & 23.77 & 0.46 & 13.37 & 23.77 \\
\hline

\end{tabular}
\end{table*}

\section{Radial intensity profiles for Models 1-11}
\begin{figure*}
    \centering
    \includegraphics[width=1.0\linewidth]{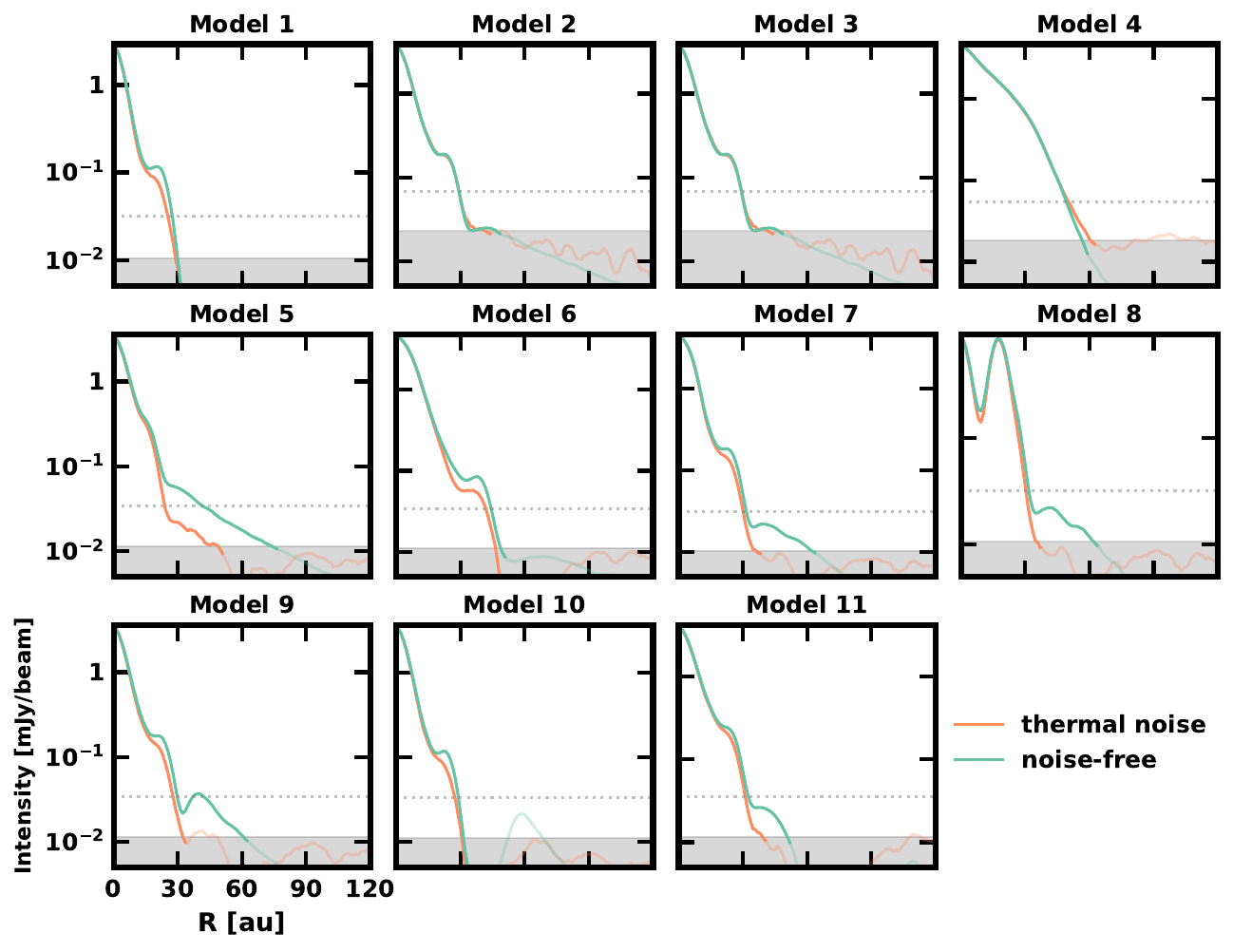}
    \caption{Azimuthally averaged radial intensity profiles of synthetic observations in ALMA Band 6 ($\lambda = 1.3$\,mm) for Models 1-11. The orange and green lines are profiles extracted from synthetic observations with and without thermal noise, respectively. The grey shades indicate the root-mean-square noise $\sigma$ measured from noisy images, and the grey dotted lines indicate the level of $3\sigma$. The synthetic observations have a beam size of $0.\arcsec05$ with an integration time of $30$ mins.\label{lastpage}}
    \label{fig:radial_intensity_all}
\end{figure*}


\bsp	
\end{document}